\documentclass[aps,prd,twocolumn,a4paper,superscriptaddress,nofootinbib]{revtex4-1}
\usepackage{graphicx}
\usepackage{multirow}
\usepackage{latexsym}
\usepackage[british]{babel}
\usepackage{soul}
\usepackage[colorlinks,linkcolor=blue,citecolor=blue,urlcolor=blue ]{hyperref}
\begin{document}

\newcommand{\be}{\begin{equation}}
\newcommand{\ee}{\end{equation}}
\newcommand{\bq}{\begin{eqnarray}}
\newcommand{\eq}{\end{eqnarray}}
\newcommand{\bsq}{\begin{subequations}}
\newcommand{\esq}{\end{subequations}}
\newcommand{\bc}{\begin{center}}
\newcommand{\ec}{\end{center}}

\title{Dark energy constraints from ESPRESSO tests of the stability of fundamental couplings}

\author{A. C. O. Leite}
\email[]{Ana.Leite@astro.up.pt}
\affiliation{Centro de Astrof\'{\i}sica, Universidade do Porto, Rua das Estrelas, 4150-762 Porto, Portugal}
\affiliation{Instituto de Astrof\'{\i}sica e Ci\^encias do Espa\c co, CAUP, Rua das Estrelas, 4150-762 Porto, Portugal}
\affiliation{Faculdade de Ci\^encias, Universidade do Porto, Rua do Campo Alegre, 4150-007 Porto, Portugal}
\author{C. J. A. P. Martins}
\email[]{Carlos.Martins@astro.up.pt}
\affiliation{Centro de Astrof\'{\i}sica, Universidade do Porto, Rua das Estrelas, 4150-762 Porto, Portugal}
\affiliation{Instituto de Astrof\'{\i}sica e Ci\^encias do Espa\c co, CAUP, Rua das Estrelas, 4150-762 Porto, Portugal}
\author{P. Molaro}
\email[]{molaro@oats.inaf.it}
\affiliation{INAF - Osservatorio Astronomico di Trieste, Via G.B. Tiepolo 11, I-34143 Trieste, Italy}
\author{D. Corre}
\email[]{david.corre@oamp.fr}
\affiliation{Laboratoire d'Astrophysique de Marseille - LAM, Universit\'{e} d'Aix-Marseille \& CNRS, UMR7326, F-13388 Marseille, France}
\author{S. Cristiani}
\email[]{cristiani@oats.inaf.it}
\affiliation{INAF - Osservatorio Astronomico di Trieste, Via G.B. Tiepolo 11, I-34143 Trieste, Italy}
\affiliation{INFN - National Institute for Nuclear Physics, via Valerio 2, I-34127 Trieste, Italy}

\date{8 July 2016}

\begin{abstract}
ESPRESSO is a high-resolution-ultra-stable spectrograph for the VLT, whose commissioning will start in 2017. One of its key science goals is to test the stability of nature's fundamental couplings with unprecedented accuracy and control of possible systematics. A total of 27 nights of the ESPRESSO Consortium's guaranteed time observations (GTO) will be spent in testing the stability of the fine-structure constant and other fundamental couplings. A set of 14 priority optimal targets have been selected for the GTO period. Here we briefly discuss the criteria underlying this selection and describe the selected targets, and then present detailed forecasts of the impact of these measurements on fundamental physics and cosmology, focusing on dark energy constraints and using future supernova type Ia surveys as a comparison point. We show how canonical reconstructions of the dark energy equation of state are improved by the extended redshift range enabled by these spectroscopic measurements, and also quantify additional improvements foreseen for a future ELT-HIRES instrument.
\end{abstract}
\maketitle

\section{Introduction}

Nature's fundamental couplings are the most mysterious part of physical theories. A fundamental constant can be defined as a parameter whose value cannot be explained within a given theory, but can only be found by measurement \cite{Uzan:2010pm}. Testing their constancy, both locally in the laboratory and in the early universe (through astrophysical observations), is therefore a powerful test of the theories where they are used: they allow us to test the domain of their validity, and if their constancy is found not to hold true to expand our knowledge by identifying clues for new physics.

ESPRESSO is the next generation high-resolution-ultra-stable spectrograph, combining the efficiency of a modern Echelle spectrograph with extreme radial velocity and spectroscopic precision, and including novel features such as improved stability thanks to a vacuum vessel and wavelength calibration done with a Laser Frequency Comb \cite{espresso}. ESPRESSO will be installed in the Combined Coud\'e Laboratory of the VLT in early 2017, and linked to the four Unit Telescopes (UT) through optical Coud\'e trains, allowing operations either with a single UT or with up to four UTs for about a 1.5 magnitude gain. The 1 UT mode will be used to achieve the highest accuracy measurements by using the brightest suitable sources while leaving the 4UT mode to perform somewhat less accurate measurements of fainter sources---which in a cosmological context typically enables access to high redshifts.

One of the key science and design drivers of ESPRESSO is to carry out improved tests of the stability of nature's fundamental couplings, and in particular to confirm or rule out the recent indications of dipole-like variations of the fine-structure constant $\alpha$ from the work of \cite{Dipole}. Such tests are of fundamental importance: a varying $\alpha$ unavoidably implies a violation of the Einstein Equivalence Principle, which in turn implies the breakdown of gravity as a purely geometric phenomenon and the presence of a fifth force in nature \cite{Uzan:2010pm}.

A minimum of ten percent of the consortium's Guaranteed Time of Observation (GTO) with the instrument will be dedicated to this purpose, which corresponds to a minimum of 27 nights if used in 1 UT mode. In order to lead to improved constraints on the stability of $\alpha$, an ideal target should present simple and strong but not saturated absorption features for the transitions with high sensitivities to such variations. In what follows we describe in more detail these criteria and apply them, together with practical constraints such as visibility from Paranal, to the full list of targets known to yield $\alpha$ measurements. We therefore arrive at a list of 14 priority targets for $\alpha$ measurements.

In addition to their intrinsic importance for fundamental physics, these tests also have a significant impact on cosmology, shedding light on the enigma of dark energy \cite{GRG}. They optimally complement traditional observables (such as Type Ia supernovas) used to map the dark energy equation of state, in particular because they significantly extend the redshift range that can be probed by the traditional methods \cite{Amendola1,Leite1}. Here we extend previous work of \cite{Leite2}, and present detailed forecasts of the impact of ESPRESSO measurements of $\alpha$ on these 14 targets. We note that these forecasts can be reliably made, once one has an accurate list of the redshift distribution of the targets, because they mostly depend on the sensitivity of the measurement rather than its central value---in other words, on the error bar, rather than on whether one has a detection or a null result. Obviously for other purposes the central value is crucial \cite{Pinho}.

In order to quantify the impact we use statistical tools previously developed for dark energy equation of state reconstructions from type Ia supernovas by \cite{HutererandStarkman}, and previously adapted for this purpose by some of the present authors in \cite{Amendola1} as well as figures-of-merit recommended by the Joint Dark Energy Mission Figure of Merit Science Working Group \cite{DEFM}. In particular, we use a 'canonical' SNAP-like future supernova type Ia survey as a comparison point. Finally we also discuss further improvements on the assumption that the same targets are observed with the European Extremely Large Telescope's high-resolution spectrograph, ELT-HIRES \cite{HIRES}, which is currently in its Phase A.

\section{Measurements of the fine-structure constant}

Quasar (QSO) absorption spectra are powerful laboratories to test the variation of dimensionless fundamental parameters such as fine structure constant, $\alpha$ and the proton-to-electron mass ratio $\mu$, as well as to test the redshift dependence of the temperature of the cosmic microwave background, $T_{CMB}$. Absorption lines produced by the intervening clouds along the line of sight of the QSO give access to physical information on the atoms/molecules present in the cloud, and this means that they give access to physics at different cosmological times and places.

The quasar spectra display metal absorption lines, which can be due to clouds either associated with the material of the galaxy that hosts the QSO, or related with some other different objects at other cosmological distances along the same line of sight. These lines are sensitive to variation of the fine structure constant, $\alpha$, and each element presents a different sensitivity to it. 
From the observational point of view the variation in $\alpha$ is defined as:
\begin{equation}
\frac{\Delta\alpha}{\alpha}= \frac{\alpha (z) -\alpha_0 }{\alpha_0}
\end{equation}
where $\alpha (z)$ is the measurement of $\alpha$ at some redshift z, and $\alpha_0$ is the laboratory value.

The Many Multiplet method \cite{Dzuba1, Dzuba2}  makes use of all transitions available in one system, each one of them with different sensitivities, $q$,  and therefore different effects in the line position on the spectra. For a given transition we can write
\begin{equation}
\Delta v \approx -\frac{2cq_{i}}{\omega_{0}} \left(\frac{\Delta\alpha}{\alpha}\right)
\end{equation}
where $\Delta v$ is the velocity shift, $c$ is the speed of light, $\omega_{0}$ is the laboratory wavelength and $q_i$ is the $q$ coefficient for a given transition.

We note that it is not the actual value of the sensitivity coefficient of a single transition $q$ that constrains $\Delta\alpha/\alpha$, since one needs to simultaneously determine the redshift of the absorber. Thus one needs to identify various transitions, with different sensitivities, all formed at the same distance from the observer. Ideally, one requires at least one transition whose wavelength decreases if $\alpha$ varies (henceforth referred to as a blue shifter), one whose wavelength increases (a red shifter) and one whose wavelength is comparatively little affected by any such variation (an 'anchor', typically from a comparatively light atom). From this it follows that the important parameter is the difference, $\Delta q$, between the largest and smallest values of the sensitivity coefficients of all the transitions available to be used in a given absorber.

\section{Targets for ESPRESSO's Fundamental Physics GTO}

\begin{figure*}
\centerline{\includegraphics[width=6.5in]{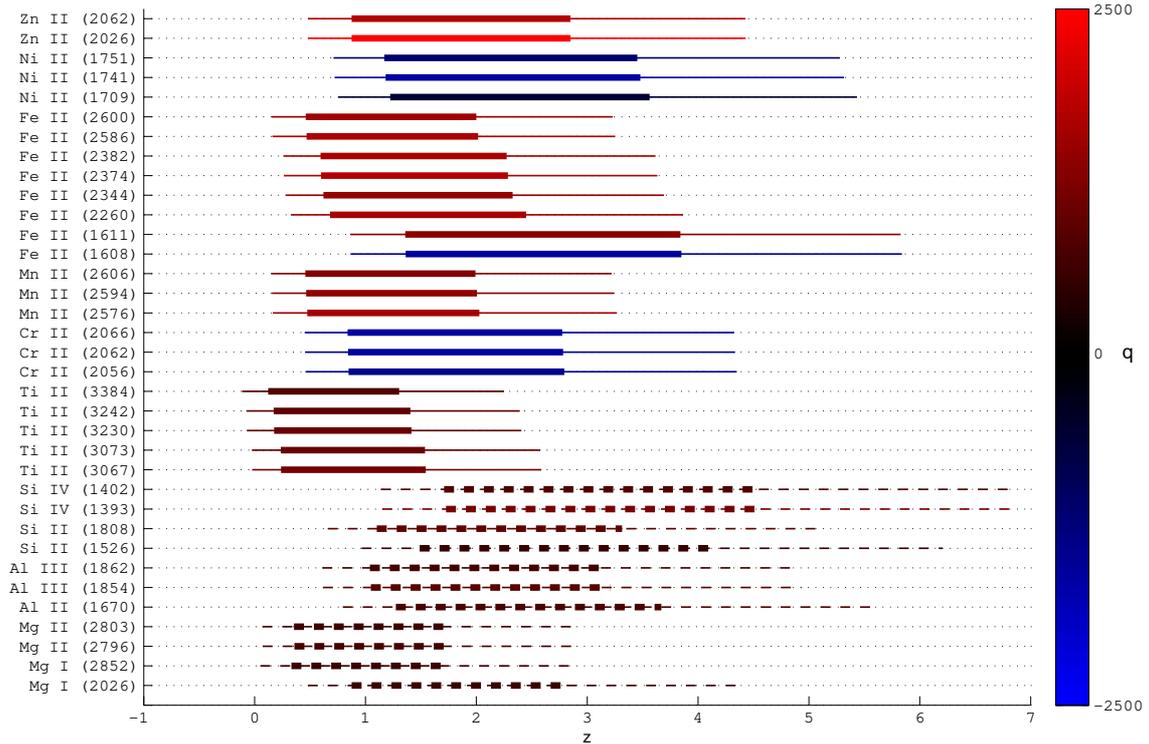}}
\caption[]{Redshift coverage of ESPRESSO and UVES of common transitions used for measuring $\alpha$. Thinner lines represent the coverage of UVES, while the thicker part is representative of ESPRESSO's. The color code is indicative of the $q$ sensitivity parameter, each transition has the color according to the blueshift or redshift on spectra and by how much. The dashed transitions at the bottom of the diagram correspond to anchors, i.e. transitions that do not shift much.}
\label{fig:transitions}
\end{figure*}

A key limitation of ESPRESSO is its wavelength coverage range, which is narrower than the ones of its predecessors (HARPS, UVES and Keck-HIRES). The effect of the shorter wavelength coverage of ESPRESSO versus the larger one from UVES is illustrated in Fig \ref{fig:transitions}. The figure depicts the redshift range in which the most common transitions used to do $\alpha$ measurements are accessible to both spectrographs. Each transition is colored by the corresponding sensitivity, $q$, to the $\alpha$ variation.

To select the list of best possible targets for the GTO of ESPRESSO we start by considering all existing measurements from the VLT-UVES and Keck-HIRES spectrographs \cite{king,murphy,chand1,chand2,chand3,lev1,lev2,lp1,lp2}, taking into account the effects the shorter wavelength coverage of the spectrograph. We chose the targets that:
\begin{itemize}
\item can be observed from the VLT site (Cerro Paranal in Chile, implying declination $< 30º$);
\item present transitions that allow a high sensitivity ( $\Delta q>2000$);
\item have a reported uncertainty of $\sigma_{\Delta\alpha/\alpha}<5ppm$.
\end{itemize} 

The last criterion comes from the fact that simple spectra should have already produced measurements with statistically lower uncertainties. Strictly speaking there is also the possibility that new bright quasars are discovered, but since the GTO targets should be fixed soon the probability of such an occurrence is low. Additional criteria that are relevant for prioritizing the targets are:
\begin{itemize}
\item QSO brightness;
\item high number of transitions available in the system, which leads to smaller overall uncertainties and also allows for several independent measurements using different sets of transitions (an important test of possible systematics);
\item presence of at least one red shifter, one blue shifter, and one anchor (as discussed in the previous section); this is partially ensured by the requirement of a large $\Delta q$;
\item simpler velocity structure systems (strong but not saturated absorption features; narrow lines and large number of components, provided these are resolved or at least partially resolved);
\item systems for which the dipole model of Webb {\it et al.} \cite{Dipole} predicts a higher variation  of $\alpha$;
\item possibility to perform in the same system additional measurements, such as $\mu$ or $T_{CMB}$, enabling key tests of many theoretical paradigms \cite{GRG}.
\end{itemize}

\begin{table*}
\caption{The best currently available measurements of $\alpha$, among the targets accessible to ESPRESSO. Column 1 gives the quasar name; the redshifts of the absorption system are given in Column 2; Columns 3 and 4 give the value of the measurement and the correspondent uncertainty. Column 5 gives the ranges of sensitivity coefficients associated with the transitions of the absorption systems. Column 6 gives the number of transitions in each absorption system and column 7 the elements that can be detected, colored differently according to whether they are an anchor (black), a blue shifter (blue) or a red shifter (red). The last column gives the references for each measurement. Measurements flagged with a * identify targets for which some of the transitions used in the current measurement are outside the wavelength range of ESPRESSO.}
\label{tb:targetalpha}
\begin{center}
\begin{tabular}{|l|c|c|c|c|c|c|c|}
\hline 	
 	Name & $z_{abs}$ & $\frac{\Delta \alpha}{\alpha}$ {\small $(10^{-6})$}  & $\sigma_{\frac{\Delta \alpha}{\alpha}} ${\small $(10^{-6})$} & Max($\Delta q$) & \# trans. & Trans. & Ref. \\
 	\hline 		 
J0350-3811 & 3.02 & -27.9 & 34.2 & 1350& 2& SiII,\textcolor{blue}{FeII} & \cite{murphy} \\ 
J0407-4410 &  2.59  & 5.7  & 3.4* & 2984& 13 & AlII,AlIII,SiII,\textcolor{blue}{CrII},\textcolor{blue}{FeII},\textcolor{red}{FeII},\textcolor{blue}{NiII},\textcolor{red}{ZnII}  & \cite{king} \\
J0431-4855 &  1.35 &  -4.0  & 2.3* & 2990& 17 & MgI,AlII,SiII,\textcolor{blue}{CrII},\textcolor{red}{MnII},\textcolor{red}{FeII},\textcolor{blue}{NiII} & \cite{king}\\
 J0530-2503 & 2.14   &6.7  &3.5* & 2990 & 7 & AlII,\textcolor{blue}{CrII},\textcolor{blue}{FeII},\textcolor{red}{FeII},\textcolor{blue}{NiII}  & \cite{king}\\
 J1103-2645 & 1.84 &5.6  &2.6  & 2890 & 4 & SiII,\textcolor{blue}{FeII},\textcolor{red}{FeII}  & \cite{lev2} \\
 J1159+0112 & 1.94   &5.1  &4.4*  & 2990 & 12 &  SiII,\textcolor{blue}{CrII},\textcolor{red}{MnII},\textcolor{blue}{FeII},\textcolor{red}{FeII},\textcolor{blue}{NiII} & \cite{king} \\
J1334+1649 & 1.77   &8.4  &4.4 &2990 & 15 &  MgII,AlII,SiII,\textcolor{blue}{CrII},\textcolor{red}{MnII},\textcolor{blue}{FeII},\textcolor{red}{FeII},\textcolor{blue}{NiII},\textcolor{red}{ZnII}& \cite{king} \\
HE1347-2457 	& 1.43  & -21.3 & 3.6   & 2790 & 3 & \textcolor{blue}{FeII},\textcolor{red}{FeII}  & \cite{lev2}\\
J2209-1944 & 1.92 & 8.5 & 3.8 &3879& 16 &AlII,SiII,\textcolor{blue}{CrII},\textcolor{red}{MnII},\textcolor{blue}{FeII},\textcolor{red}{FeII},\textcolor{blue}{NiII},\textcolor{red}{ZnII}  & \cite{king} \\
HE2217-2818 & 	1.69     & 1.3 & 2.4  &2890 & 6 & AlIII,\textcolor{blue}{FeII},\textcolor{red}{FeII} & \cite{lp1}\\
Q2230+0232 & 	1.86    & -9.9 & 4.9 &3879 & 14 & SiII,\textcolor{blue}{CrII},\textcolor{blue}{FeII},\textcolor{red}{FeII},\textcolor{blue}{NiII},\textcolor{red}{ZnII} & \cite{murphy} \\
J2335-0908 & 2.15  & 5.2 & 4.3* &3879& 16 &  AlIII,\textcolor{blue}{CrII},\textcolor{blue}{FeII},\textcolor{red}{FeII},\textcolor{blue}{NiII},\textcolor{red}{ZnII}  & \cite{king}\\
J2335-0908 & 2.28  & 7.5 & 3.7*   &2610& 7 &  SiIV,\textcolor{blue}{CrII},\textcolor{blue}{FeII},\textcolor{red}{FeII},\textcolor{blue}{NiII}  & \cite{king}\\
Q2343+1232 & 	2.43     & -12.2 & 3.8* & 3879 & 11 & AlII,SiII,\textcolor{blue}{CrII},\textcolor{blue}{FeII}   \textcolor{blue}{NiII},\textcolor{red}{ZnII}& \cite{murphy}\\
\hline 
\end{tabular}
\end{center}
\end{table*}

This analysis leads to the selection of the 14 targets which are presented in Table \ref{tb:targetalpha}. We note that the order in which they are presented should not be seen as any ranking among them: they are simply ordered according to their Right Ascension. A more detailed prioritization will require the generation of simulated ESPRESSO-like spectra of these targets, and is currently ongoing.

Strictly speaking, the first listed measurement does not fulfill all the criteria, but it is the only system accessible to ESPRESSO where the proton-to-electron mass ratio and the temperature-redshift relation can also be measured. This fact makes it a theoretically interesting target for testing theories where a relation between these three parameters is predicted \cite{ferreira1,ferreira2}. Indeed, this target has also yielded a measurement of the Deuterium abundance \cite{Deuterium}, but unfortunately the required transitions (having rest wavelengths around 915 \AA) are not accessible to ESPRESSO at this redshift.

\begin{figure}
  \centering
  \begin{tabular}{c}
  
    \includegraphics[width=1\linewidth]{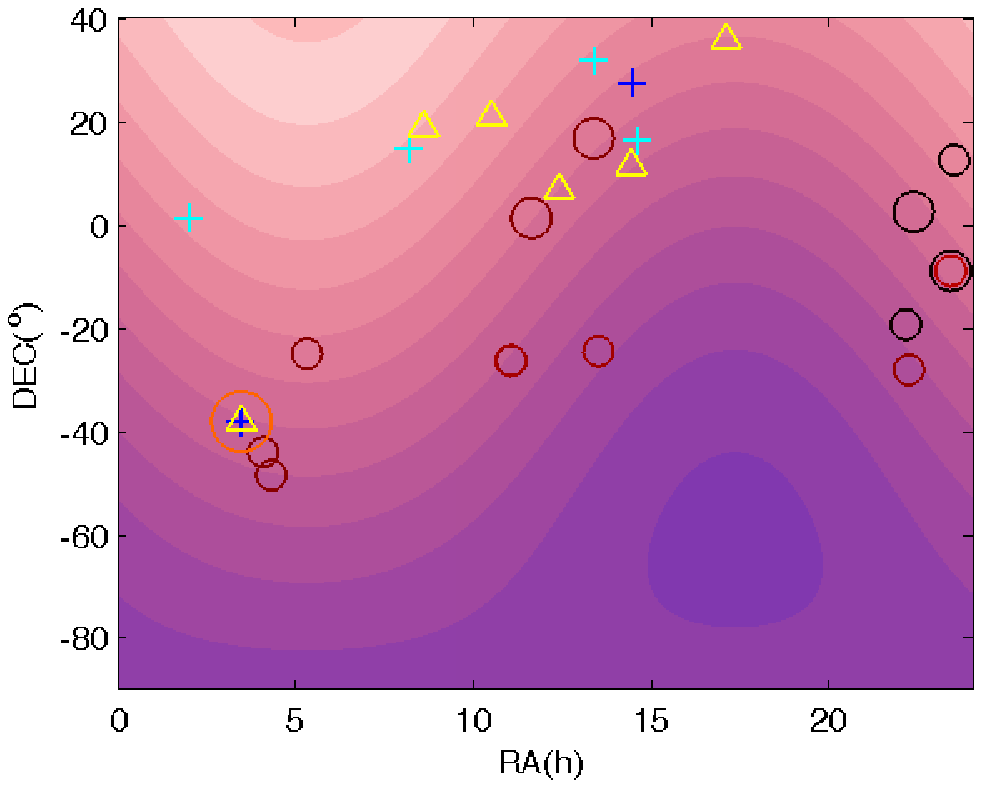}\\
    \includegraphics[width=1\linewidth]{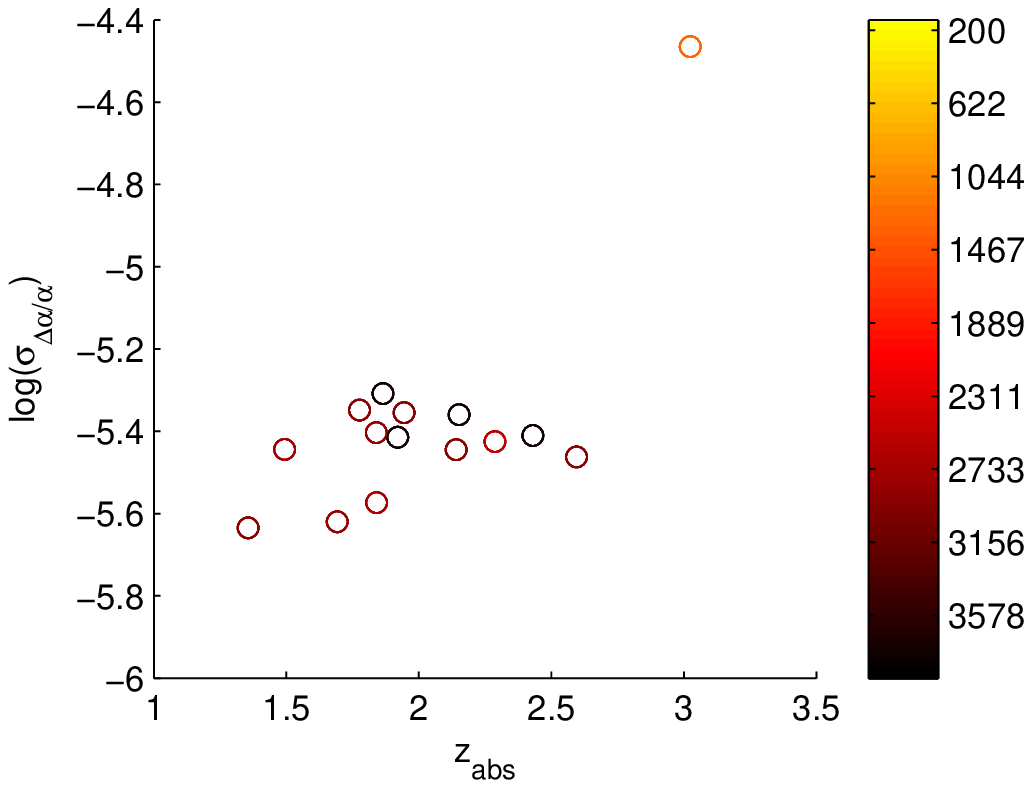}\\
  \end{tabular}
  \caption{\label{fig1obsmap} The top panel shows the sky distribution of the ESPRESSO targets in in Table \protect\ref{tb:targetalpha}. The circles represent the $\alpha$ measurements, their size being proportional to the uncertainty of the current measurement and the color gives the value for $\Delta q$ according to the color bar in the lower panel. The dark blue crosses represent known $\mu$ measurements, and the cyan possible targets of $\mu$ (for which no measurement currently exists). The yellow triangles represent targets for $T_{CMB}$ measurements. The  plot's colormap represents the expected value of $\alpha$ in each direction according to the best-fit dipole prediction for the value of $\alpha$ \protect\cite{Dipole}, with the purple representing a positive shift and the light pink a negative shift. The bottom panel shows the redshift and uncertainty of the current $\alpha$ measurements, the color corresponding to the value for $\Delta q$ according to the values of the color bar.}
\end{figure}

Figure \ref{fig1obsmap} presents, in a visually simpler way, the information listed in Table \ref{tb:targetalpha}, considering the sky position and the redshift of the targets. In particular, these plots facilitate a rapid visualization of the target position to test the dipole model of Webb {\it et al.}, represented by the color map in the background, and the target distribution in redshift, which ranges from 1.35 to 3.02 (or 2.59 if not considering the first target in the table). In either case, we note that all these absorbers probe physics deep in the matter era, an epoch which is quite difficult to probe with more common probes such as Type Ia supernovas. This is important because, as previously mentioned, even improved null measurements of $\alpha$ can be used to constrain dark energy, a point to which we now turn.

\section{Forecasts on dark energy}

Our Principal Component Analysis (PCA) based formalism to obtain the forecasts is described in \cite{Amendola1}. For completeness, a short review of the method is presented in Appendix \ref{App:AppendixA_PCA}. We consider models for which the variation of $\alpha$ is linearly proportional to the displacement of a scalar field, and further assume that this field is a quintessence type field, i.e. responsible for the current acceleration of the Universe.

We take the coupling between the scalar field and electromagnetism to be:
\begin{equation}
{\cal L}_{\phi F} = - \frac{1}{4} B_F(\phi) F_{\mu\nu}F^{\mu\nu} ,
\end{equation}
where the gauge kinetic function is $B_{F}(\phi)=1-\zeta\kappa(\phi-\phi_{0})$, $\kappa^{2}=8\pi G $ and $\zeta$ is a dimensionless parameter to be marginalized over. This can be seen as the first term of a Taylor expansion, and should be a good approximation if the field is slowly varying at low redshift---a good approximation given that it is assumed to be a quintessence-type field and that observationally only small relative variations of $\alpha$ are allowed \cite{Carroll,Dvali,Chiba}. Then, the evolution of $\alpha$ is given by
\begin{equation} 
\frac{\Delta \alpha}{\alpha} \equiv 
\frac{\alpha-\alpha_0}{\alpha_0} = \zeta \kappa (\phi-\phi_0) \,. 
\end{equation}
For a flat Friedmann-Lema\"{i}tre-Robertson-Walker Universe with a canonical scalar field we can write its speed as $\dot{\phi}^{2}=(1+w(z))\rho_{\phi}$, from which it follows that for a given dependence of the equation of state parameter $w(z)$ with redshift the scalar field evolves as
\begin{equation} 
\phi(z)-\phi_0 = \frac{\sqrt{3}}{\kappa} 
\int_0^z \sqrt{1+w(z)} \left(1+ \frac{\rho_m}{\rho_\phi}\right)^{-1/2} 
\frac{dz}{1+z} . 
\end{equation} 
where we have chosen the positive root of the solution since we expect the scalar field to be rolling down the potential.

From this one can calculate the Fisher matrix  to infer the precision on the measurement of $w$ using standard techniques \cite{Amendola1} obtaining, from a set of observables and its uncertainties, the eigenvalues $\lambda_{i}$ of the diagonalized Fisher matrix (ordered from best determined modes to worst ones) and the variance of the new parameters, $\sigma_{i}^{2}=1/\lambda_{i}$. To reconstruct the fiducial equation only the best determined modes are used, chosen according to statistical and physical criteria detailed in \cite{Amendola1}.

In this work we will consider three fiducial models for the equation of state. First we will assume a constant equation of state, with a value close to a cosmological constant, $w_c(z)=-0.9$. This has already been used in previous works, and therefore using it facilitates comparisons of our results with those of previous works. 

Second, we consider a Chevallier-Polarski-Linder \cite{CPL1,CPL2} parametrization of the dark energy equation of state,
\begin{equation} \label{cpl}
w_{\rm CPL}(z)=w_0+w_a \frac{z}{1+z}\,,
\end{equation}
where $w_0$ is its present value and $w_a$ is the coefficient of the time-dependent term. We have chosen the fiducial parameters $w_0=-0.9$ and $w_a=0.3$.

Third, we consider an Early Dark Energy (EDE) class of models \cite{EDE} for which  the dark energy density fraction is
\begin{equation}
\Omega_{\rm EDE}(z) = \frac{1-\Omega_m - \Omega_e \left[1- (1+z)^{3 w_0}\right] }{1-\Omega_m + \Omega_m (1+z)^{-3w_0}} + \Omega_e \left[1- (1+z)^{3 w_0}\right] \label{edeomega} 
\end{equation}
and the  dark energy equation of state is defined as
\begin{equation}
w_{\rm EDE}(z)=-\frac{1}{3[1-\Omega_{\rm EDE}]} \frac{d\ln\Omega_{\rm EDE}}{d\ln a} + \frac{a_{eq}}{3(a + a_{eq})}\,;
\label{eq:edew}
\end{equation}
here $a_{eq}$ is the scale factor. The energy density $\Omega_{\rm EDE}(z)$ has a scaling behavior evolving with time and approaching a finite constant $\Omega_e$ in the past, rather than approaching zero. Here used $w_0=-0.9$, $\Omega_e=0.02$ and $z_{eq}=3371$. Therefore the second and third parametrizations are extensions of the first one, each of them having one additional parameter.

For each fiducial model we choose a prior for the coupling parameter $\zeta=5\times10^{-6}$ such that it leads to a few parts-per-million variation of $\alpha$ at redshift $z\sim4$, consistent with \cite{Dipole} and with other current data.

We applied this PCA formalism to the dataset of 14 ESPRESSO $\alpha$ targets discussed above, on its own and also in combination with representative future Sna Ia Surveys. For the $\alpha$ measurements we assumed two different scenarios for the ESPRESSO GTO target list: 
\begin{itemize}
\item \textbf{Baseline}: we assumed that each of the targets on the list can be measured by ESPRESSO with an uncertainty of $\sigma_{\Delta\alpha / \alpha}=0.6ppm$; this represents what we can currently expect to achieve (though this expectation needs to be confirmed at the time of commissioning); 
\item \textbf{Ideal}: in this case we assumed a factor of 3 improvement in the uncertainty, $\sigma_{\Delta\alpha / \alpha}=0.2ppm$; this represents somewhat optimistic uncertainties, but is also useful as a comparison point.  
\end{itemize}
We will also provide forecasts for a longer-term dataset, on the assumption that the same targets are observed with the ELT-HIRES spectrograph; in this case we assume an improvement in sensitivity by a factor of 6 in both the baseline and ideal scenarios, coming from the larger collecting area of the telescope and additional improvements at the level of the spectrograph.

As for the Type Ia supernovas, we consider the following datasets: 
\begin{itemize}
\item  A low-redshift sample, henceforth denoted \textbf{LOW}, of 3000 supernovas uniformly distributed in the redshift range $0 < z < 1.7$, with an uncertainty on the magnitude of $\sigma_{m} = 0.11$. These numbers are typical of a `SNAP-like' future supernova survey and were also used in \cite{HutererandStarkman} and many other subsequent works, thus providing a useful point of comparison;
\item An intermediate redshift sample, henceforth denoted \textbf{MID}, of 1700 supernovas uniformly
distributed in the redshift range $0.75 < z < 1.5$ and the same $\sigma_{m}$ as before.
This is representative of recent proposals such as DESIRE \cite{Astier}
\end{itemize}

To reconstruct the dark energy equation of state, we assumed 20 redshift bins between $0 < z < 3.02$, though note that the ESPRESSO GTO $\alpha$ measurements only occupy the range $1.35<z<3.02$, and similarly the type Ia supernova data only cover comparatively low redshifts. Increasing the redshift coverage is indeed one of the key advantages of these $\alpha$ measurements, as discussed in \cite{Amendola1}. In order to compare the reconstructions with the different datasets, and thus to quantify the gains in sensitivity from the combination of several of them, we compared the uncertainty of the five best-determined PCA modes of the various reconstructions. We take the reconstruction with the \textbf{LOW} Supernova data as the baseline, and compare the others with this one by taking the ratios
 \begin{equation}\label{eq:ratio}
 {\cal R}_{i,j} =  \frac{\sigma_{i,j}^{-2}}{\sigma_{LOW,j}^{-2}}
 \end{equation}
in which $i$ is the dataset (or combination thereof) under consideration and $j$, which ranges from 1 to 5, denotes the PCA mode being considered. The mode uncertainties used should be normalized in each reconstruction. One expects that the more accurate reconstructions will result in larger values of this ratio. This procedure is detailed in \cite{DEFM}, which also suggests the graphs of the principal components as a function of redshift as means to provide information on the redshift coverage and sensitivity of the reconstruction.

\begin{figure*}
\begin{center}
\includegraphics[width=6in]{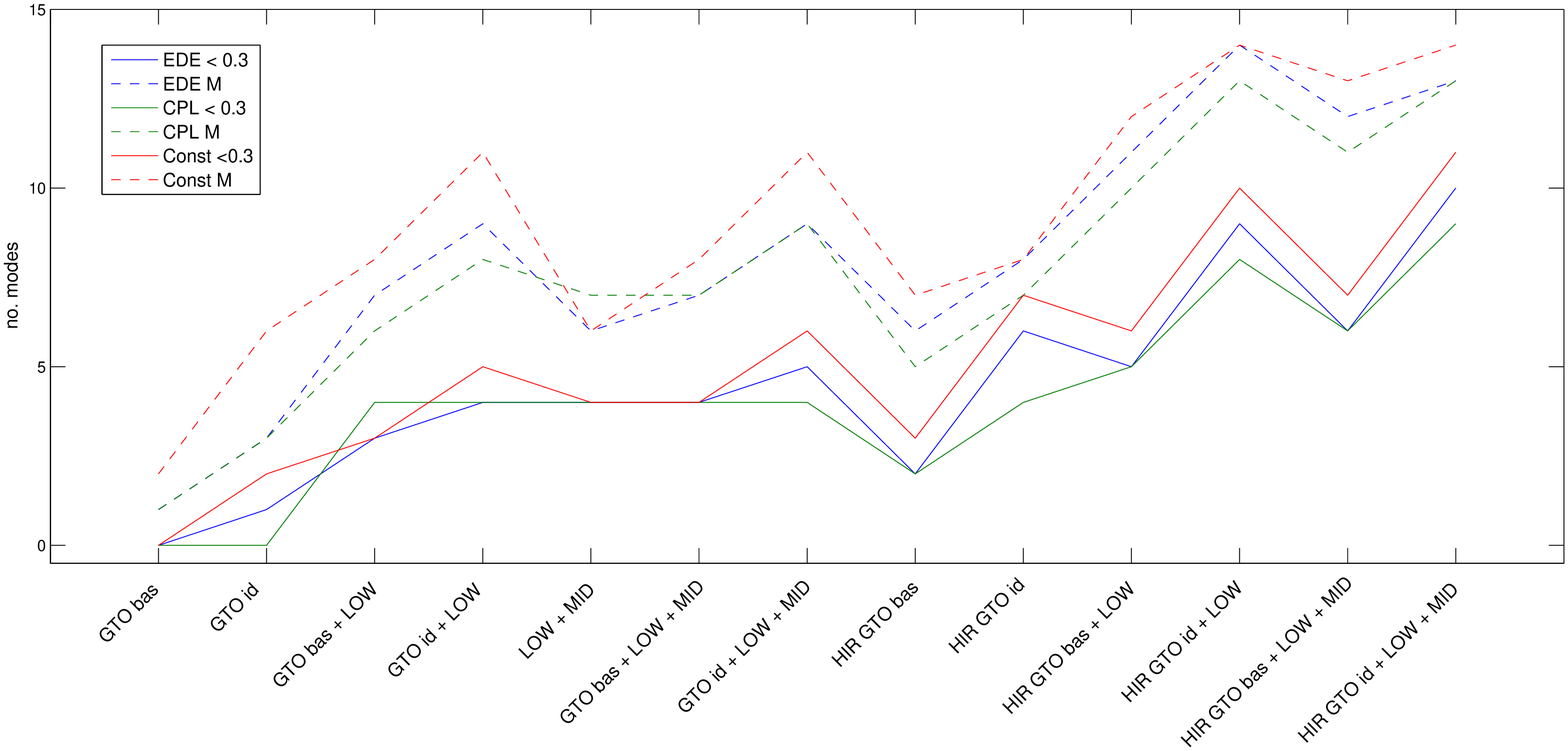}
\end{center}
\caption{The number of PCA modes with uncertainties smaller than 0.3 (solid lines) and the optimal number of PCA modes for the equation of state reconstruction (c.f.  Appendix \protect\ref{App:AppendixA_PCA}, dashed lines), for all the combinations of fiducial models and datasets we study. The red, green and blue lines correspond to the constant equation of state, CPL and EDE parametrizations respectively.}
\label{figcomparing}
\end{figure*}

To facilitate comparisons with previous works, we will also consider two other indicators to characterize the reconstructions: the number of modes with uncertainty lower that 0.3 and the optimal number of modes to use in the PCA reconstruction of the equation of state, as chosen using the so-called risk minimization method (as explained in in appendix \ref{App:AppendixA_PCA}). The latter number is larger than the former but the two are correlated, as can be seen in Fig. \ref{figcomparing}.

\begin{table*}
  \begin{center}
  \caption{The values of the ratios defined in Eq. \ref{eq:ratio} for the first five modes, plus the two additional diagnostics, considering reconstructions for the  \textbf{constant} fiducial equation of state parametrization. We present different combinations of datasets for baseline and ideal scenarios for measurements of $\alpha$ expected for ESPRESSO GTO and the LOW and MID type Ia supernova samples, using the LOW supernova sample as the benchmark. The bottom half of the table shows analogous forecasts on the assumption that the same $\alpha$ targets are observed with the ELT-HIRES.}
		\begin{tabular}{|l|c|c|c|c|c|c|c|}		
		\hline
				 & 1 & 2 & 3 & 4 & 5 & $<0.3$ & M \\
          \hline
    GTO baseline   & 0.0062   &  0.0095 &    0.0511  &   0.1144    & 0.3294 & 0 & 2\\      
     GTO ideal   &  0.0444   &  0.0564   &  0.1758   &  0.2773   &  0.6193 & 2 & 6\\ 
    \hline
	    GTO baseline + LOW   &   1.0000  &   1.0000   &  1.0001  &   1.0019  &   1.3173 & 3& 8\\
    GTO ideal + LOW   &  1.0000  &   1.0000   &  1.1421   &  2.6443   &  4.6373 & 5& 11\\
    \hline
    LOW + MID & 2.1318   &  2.1292  &   2.0930   &  2.0268   &  1.8011& 4 & 6\\ 
    GTO baseline + LOW + MID  & 2.1318   &  2.1292  &   2.0932   &  2.0284  &   1.8271 &4 & 8\\
    GTO ideal + LOW + MID  & 2.1318   &  2.1292  &   2.0945   &  3.0226  &   6.3953  &6 & 11\\
\hline\hline
    HIRES GTO baseline& 0.1735 &   0.2149  &  0.5966 &   0.8272 &   1.5976& 3 & 7\\
    HIRES GTO ideal   & 1.9136 &   2.3506  &  6.2696 &   8.2402 &  14.7864& 7 & 8\\
    HIRES GTO baseline + LOW   & 1.0000 &   1.0001  &  4.4515 &   5.5921 &   8.2380& 6 & 12  \\
    HIRES GTO ideal + LOW    & 1.9137  &  2.6166  & 23.0562 &  25.9220 &  51.7019& 10 & 14\\
    HIRES GTO baseline + LOW + MID   & 2.1318  &  2.1293  &  4.4538  &  5.7596 &  16.8022 & 7 & 13 \\
    HIRES GTO ideal + LOW + MID  &2.1319  &  5.0064 &  23.1224  & 55.0540&   51.7173& 11 &14\\
\hline         
         \end{tabular}
         \label{tab:recwconst}
         \end{center}
\end{table*}%

\begin{table*}
  \begin{center}
  \caption{Same as Table \protect\ref{tab:recwconst} for the \textbf{CPL} fiducial equation of state parametrization. $w_{0}=-0.9$, $w_{a}=0.3$.}
	\begin{tabular}{|l|c|c|c|c|c|c|c|}		
		\hline
				 & 1 & 2 & 3 & 4 & 5 & $<0.3$ & M \\
		 \hline
    GTO baseline   &0.0021 &   0.0048 &   0.0329 &   0.0885 &   0.2662 & 0 &1\\      
     GTO ideal   &   0.0091 &   0.0137 &   0.0548  &  0.1166 &   0.3182 & 0 & 3  \\ 
    \hline	
    GTO baseline + LOW   & 1.0000 &   1.0000 &   1.0000 &   1.0004 &   1.0042 & 4 & 6\\
    GTO ideal + LOW   &  1.0000 &   1.0000 &   1.0003  &  1.0055 &   1.9969  & 4 & 8\\
    \hline
    LOW + MID &2.1320 &   2.1292 &   2.0991  &  2.0370 &   1.8390 & 4 & 7\\ 
    GTO baseline + LOW + MID  &  2.1320 &   2.1292 &   2.0991 &   2.0335 &   1.8428& 4 & 7 \\
    GTO ideal + LOW + MID  & 2.1320 &   2.1292 &  2.0994 &   2.0411 &   2.0749 & 4 &9  \\
\hline\hline
    HIRES GTO baseline&  0.0327  &  0.0439&    0.1286 &   0.2113&    0.4935 & 2 & 5\\
    HIRES GTO ideal &0.3517 &   0.4508 &   1.1239  &  1.4889 &    2.8578 &4 & 7\\
    HIRES GTO baseline + LOW&1.0000 &   1.0000 &   1.0029  &  2.3508&    3.3813 &5 &10\\
    HIRES GTO ideal + LOW&  1.0000 &   0.9994 &   8.9818  & 10.5674&   9.7010 & 8 & 13\\
    HIRES GTO baseline + LOW + MID& 2.1320 &   2.1292 &   2.1006  &  2.3783&    6.2142 & 6 & 11\\
    HIRES GTO ideal + LOW + MID& 2.1320 &   2.1297 &  8.9749  & 10.5291&   18.0015& 9 & 13\\    
    \hline 
         \end{tabular}%
         \label{tab:recwCPL}
         \end{center}
\end{table*}%

\begin{table*}
  \begin{center}
  \caption{Same as Table \protect\ref{tab:recwconst} for the \textbf{EDE} fiducial equation of state parametrization.  $w_{0}=-0.9$, $\Omega_{EDE}=0.02$.}
\begin{tabular}{|l|c|c|c|c|c|c|c|}		
		\hline
				 & 1 & 2 & 3 & 4 & 5 & $<0.3$ & M \\
	    \hline
    GTO baseline   &0.0028 &   0.0056 &   0.0381 &   0.0976 &   0.2929 &0 & 1\\      
     GTO ideal   &   0.0149 &   0.0216 &   0.0731  &  0.1515 &   0.3846 & 1 & 3   \\ 
    \hline	
    GTO baseline + LOW   & 1.0000 &   1.0000 &   1.0000 &   1.0007 &   1.0083 & 3&  7\\
    GTO ideal + LOW   &  1.0000 &   1.0000 &   1.0006  &  1.0553 &   3.0416 & 4 & 9 \\
    \hline
    LOW + MID &2.1318 &   2.1292 &   2.0952  &  2.0303 &   1.8144 & 4 & 6\\ 
    GTO baseline + LOW + MID  &  2.1318 &   2.1292 &   2.0952 &   2.0310 &   1.8213 &4 &7 \\
    GTO ideal + LOW + MID  &  2.1318 &   2.1292 &   2.0956 &   2.0381 &   3.2039 &5 &9  \\
\hline\hline
    HIRES GTO baseline&  0.0557  &  0.0757&    0.1915 &   0.3333&    0.6940 & 2 &6\\
    HIRES GTO ideal &0.6051 &   0.8039 &   1.7870  &  2.7848&    4.8660 &6&8\\
    HIRES GTO baseline + LOW&1.0000 &   1.0000 &   1.4234  &  2.7059&    5.9279 &5 & 11\\
    HIRES GTO ideal + LOW&  1.0000 &   1.6629 &   9.3019  & 20.1404&   14.9466&9&14\\
    HIRES GTO baseline + LOW + MID& 2.1318 &   2.1293 &   2.0982  &  3.8437&    6.6324& 6 &12\\
    HIRES GTO ideal + LOW + MID& 2.1319 &   2.1302 &  15.4443  & 20.1964&   17.5422& 10 &13\\    
    \hline 
         \end{tabular}%
         \label{tab:recwede}
         \end{center}
\end{table*}%

Tables \ref{tab:recwconst}, \ref{tab:recwCPL} and \ref{tab:recwede} present the values of these ratios given by Eq. \ref{eq:ratio} for the five best determined modes and the various combinations of datasets, as well as the two other diagnostics described in the previous paragraph, for each of the fiducial models. One finds that on its own, the reconstruction with the ESPRESSO Target List for the GTO is not competitive with the one from a large Supernova dataset. This is expected given the earlier analysis in \cite{Leite2}.

\begin{figure*}
\begin{center}
\includegraphics[width=3.3in]{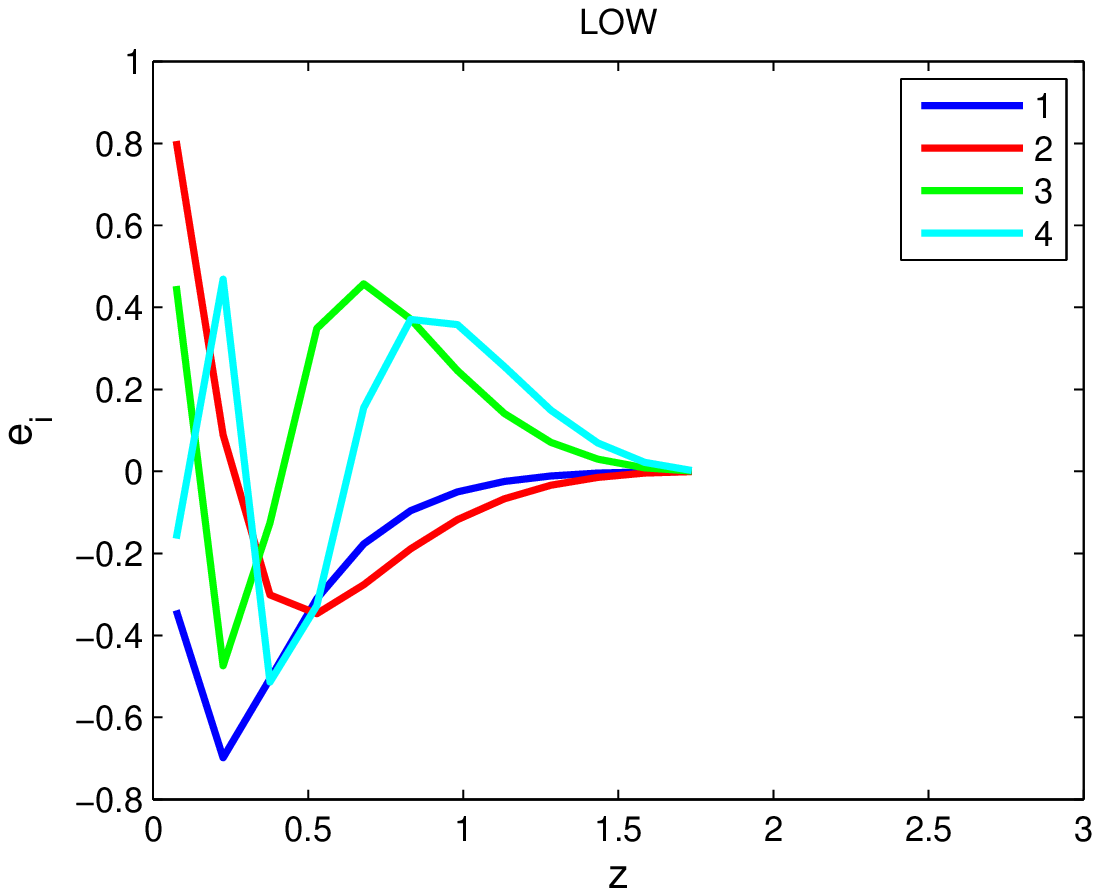}
\includegraphics[width=3.3in]{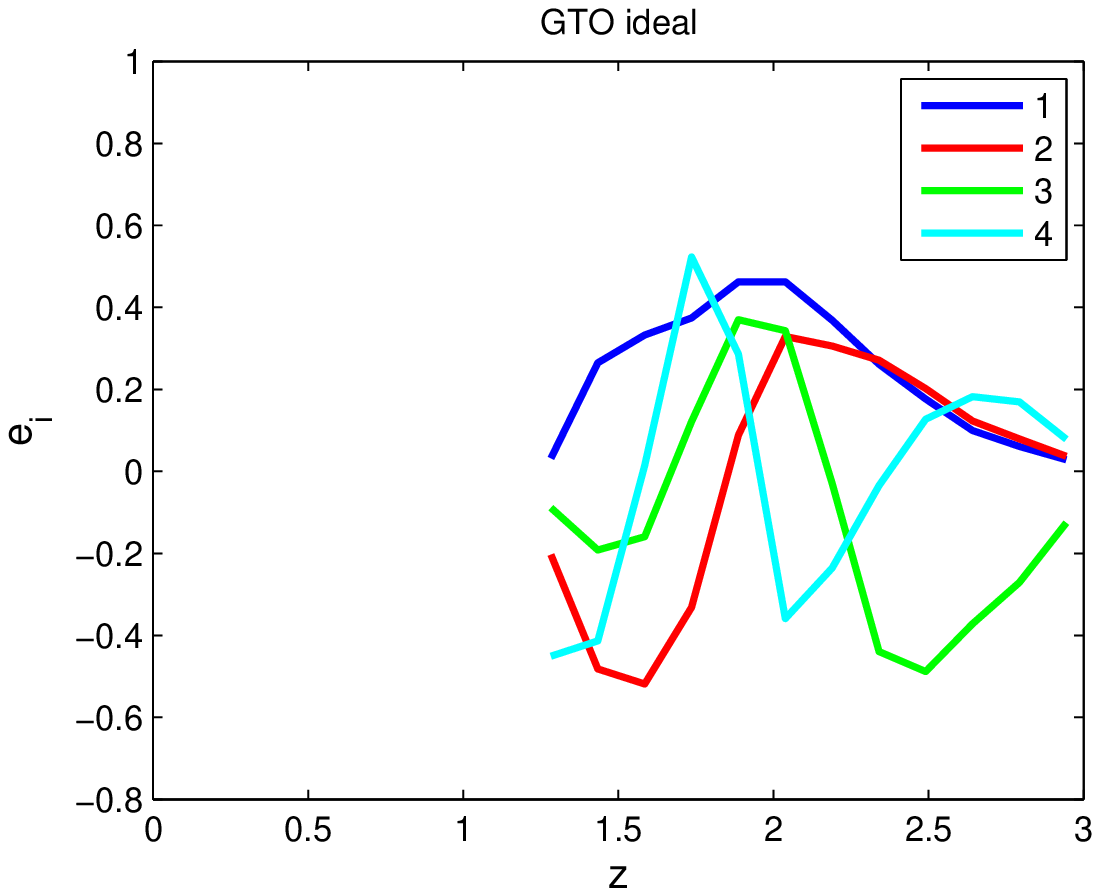}
\includegraphics[width=3.3in]{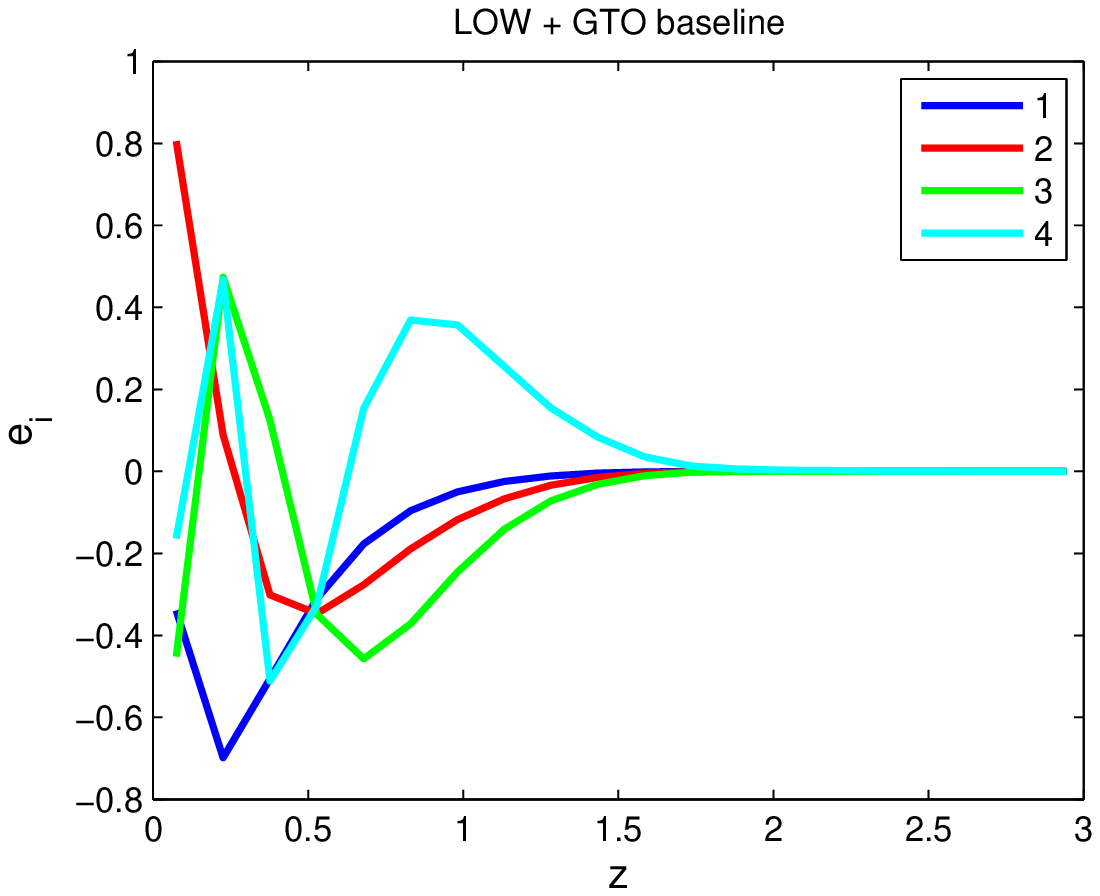}
\includegraphics[width=3.3in]{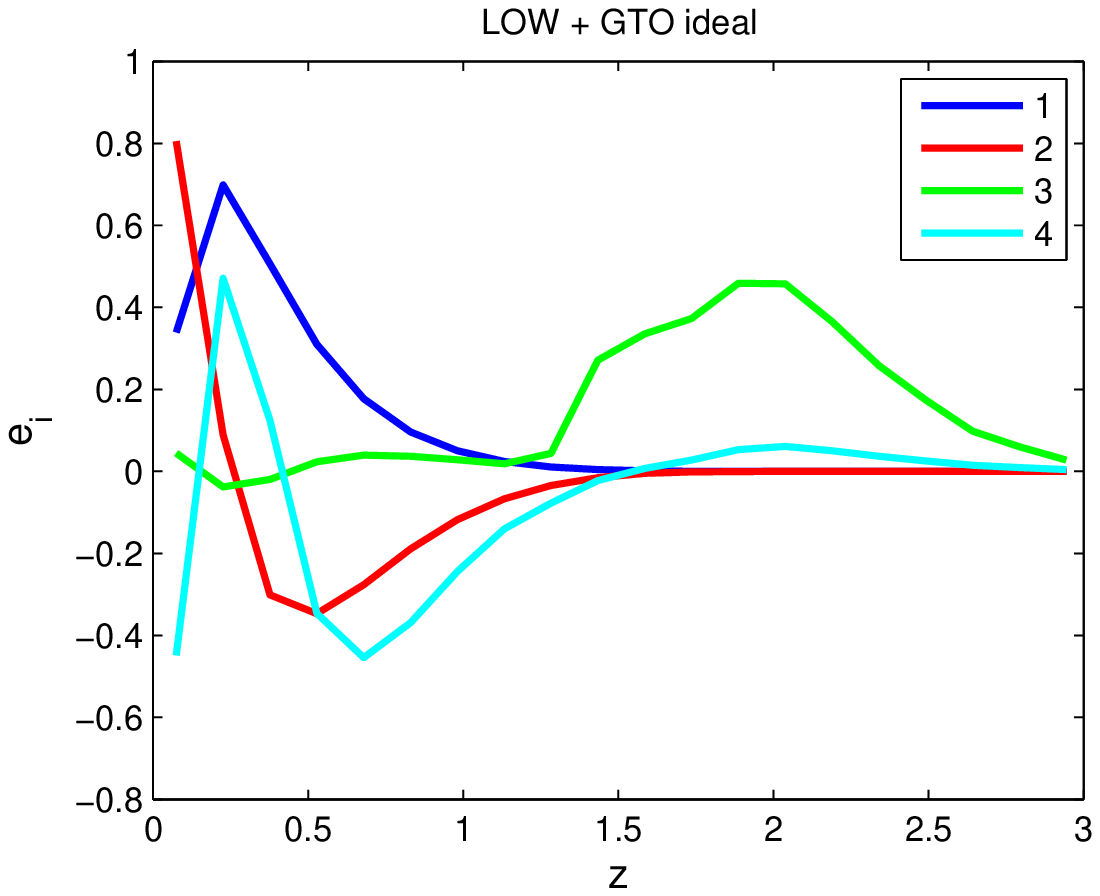}
\includegraphics[width=3.3in]{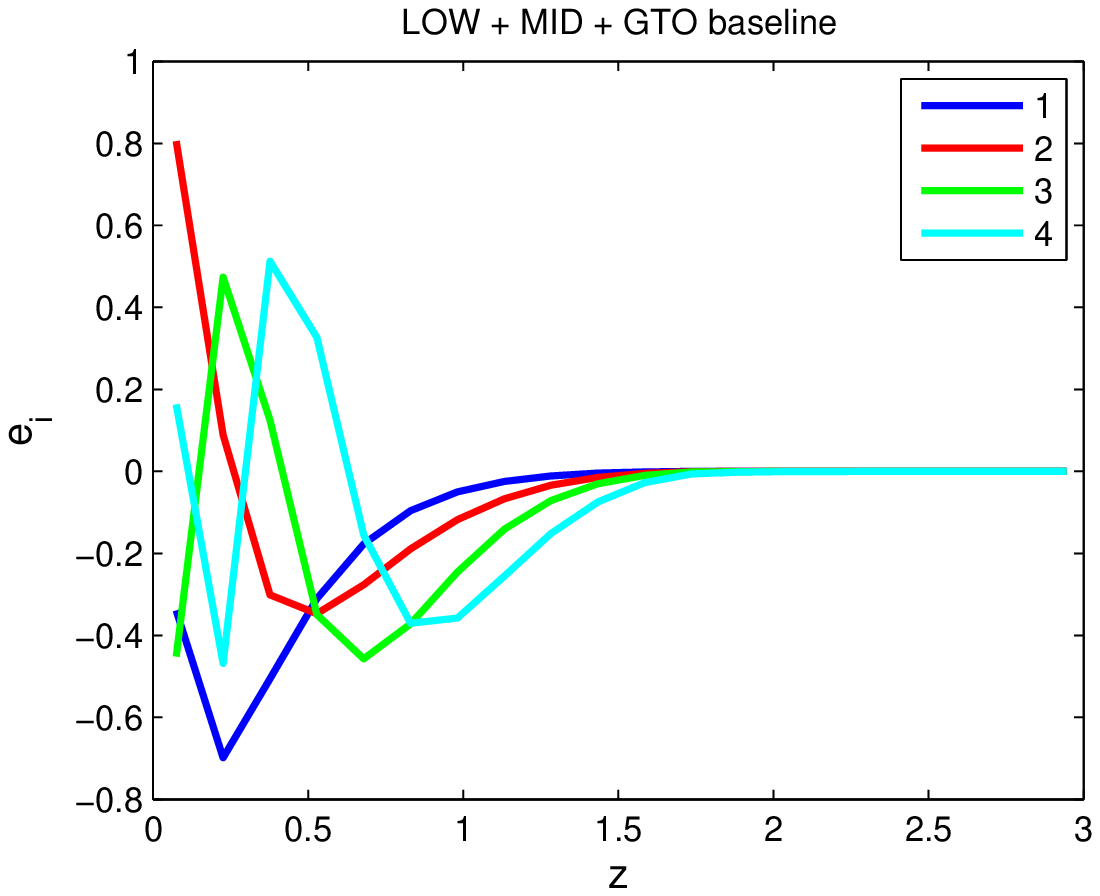}
\includegraphics[width=3.3in]{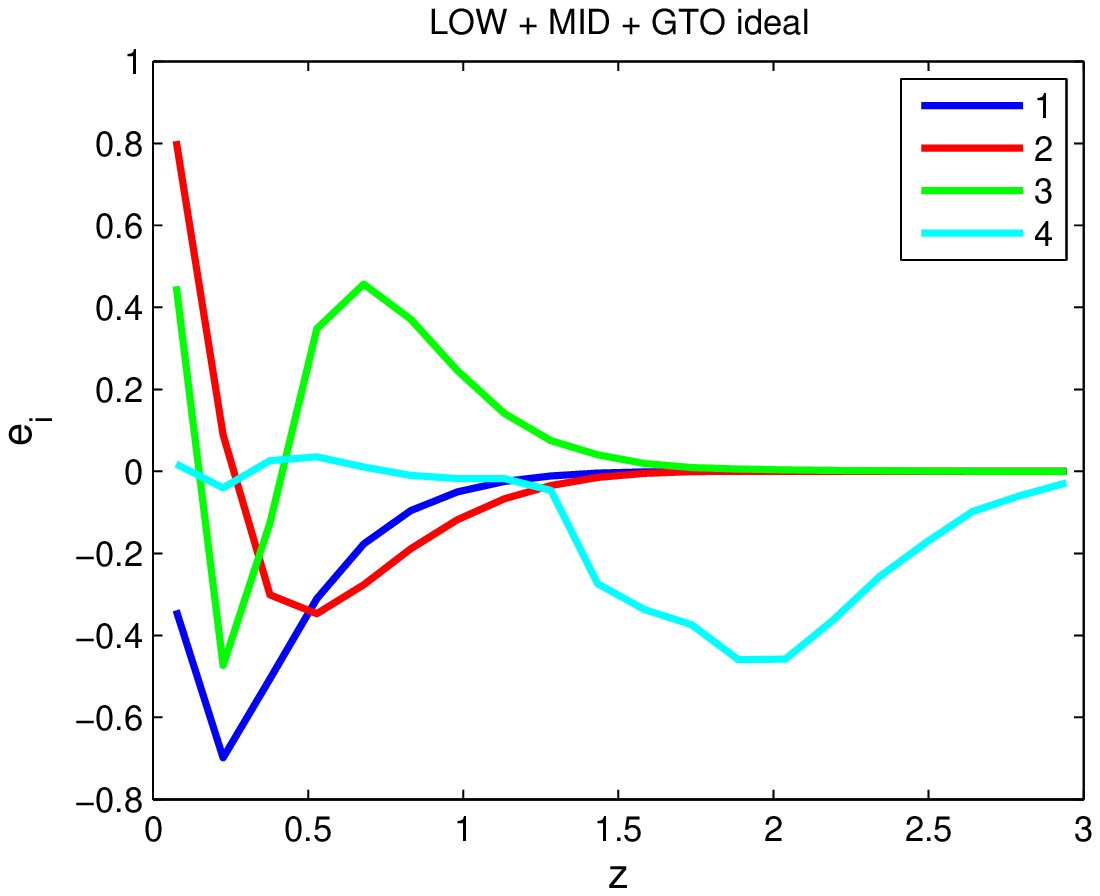}
\end{center}
\caption{Comparing the four best determined eigenmodes for the constant equation of state parametrization, for the following cases: supernovas only (LOW sample, top left), $\alpha$ measurements only (ESPRESSO GTO ideal sample, top right), LOW plus ESPRESSO GTO Baseline (middle left), LOW plus ESPRESSO GTO Ideal (middle right), LOW plus MID plus ESPRESSO GTO Baseline (bottom left), and LOW plus MID plus ESPRESSO GTO Ideal (bottom right). Note that the redshift coverage in the top panels differs from that in the remaining ones. All eigenvectors are similarly normalized; how much they differ from zero (in absolute value) indicates their sensitivity to different redshifts.}
\label{figmodes}
\end{figure*}

\begin{figure*}
\begin{center}
\includegraphics[width=3.3in]{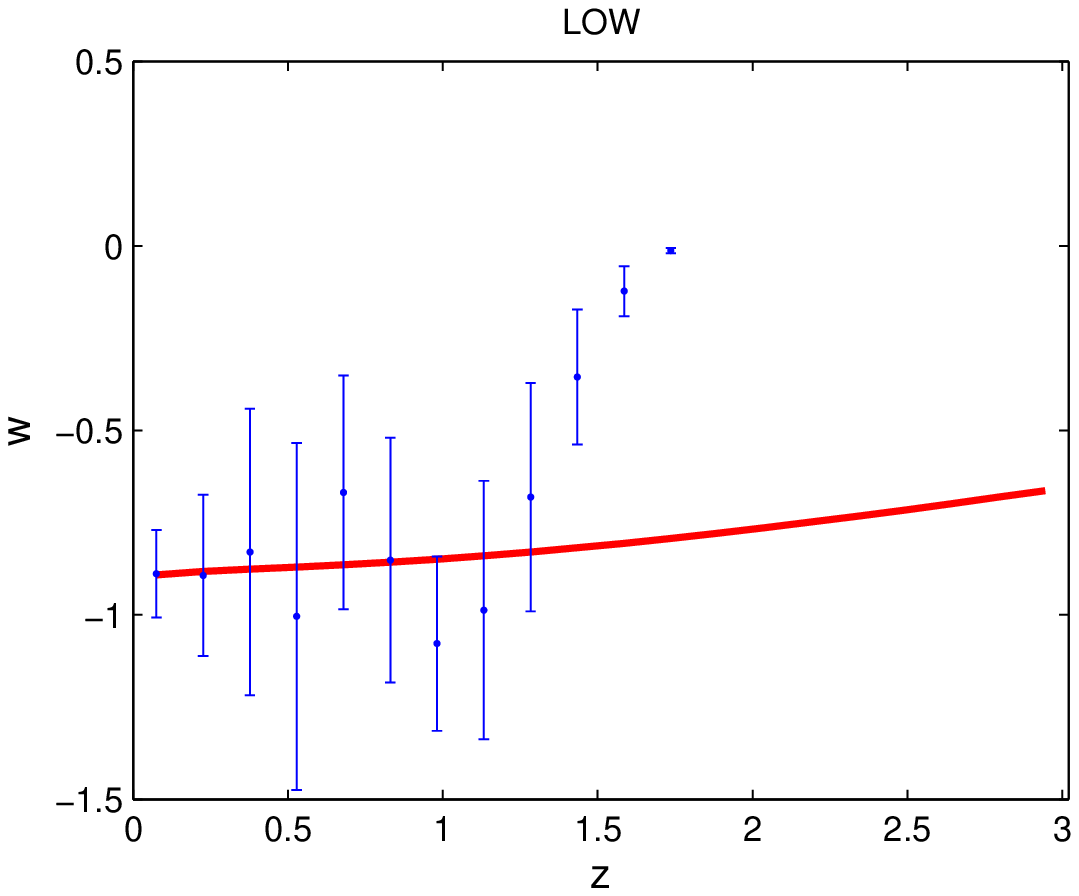}
\includegraphics[width=3.3in]{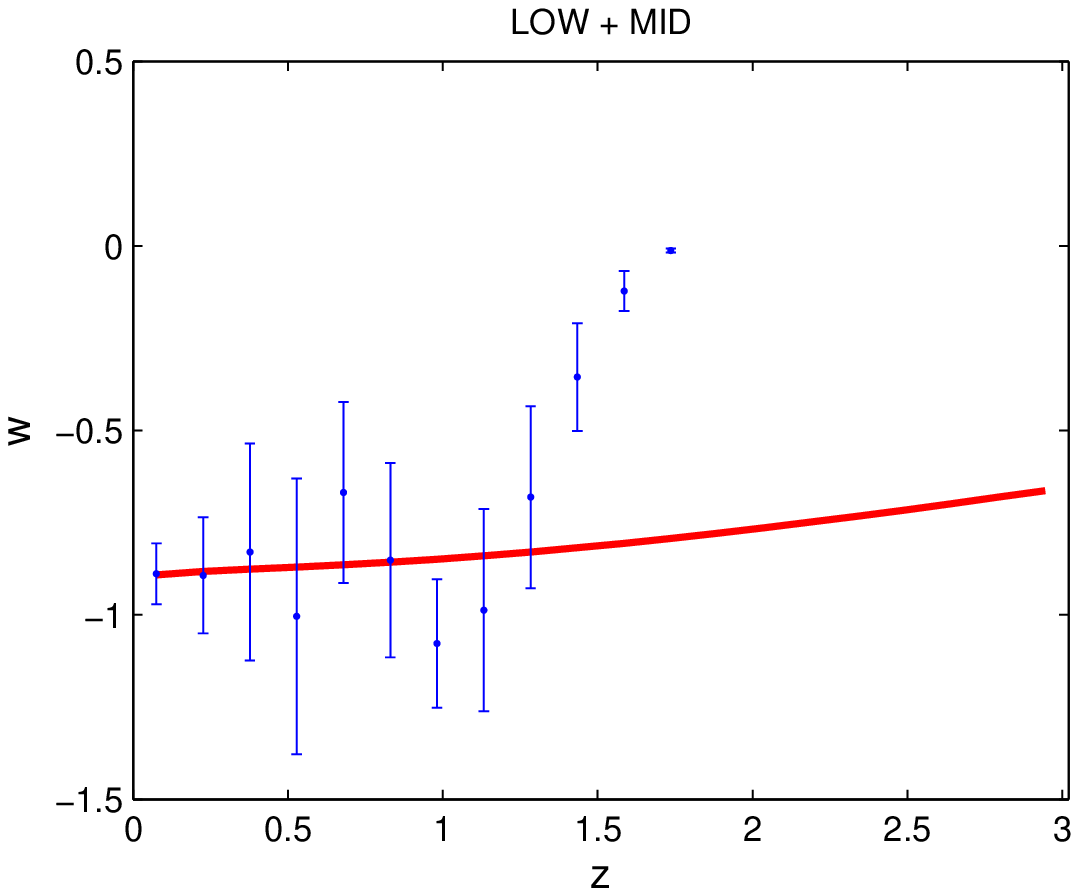}
\includegraphics[width=3.3in]{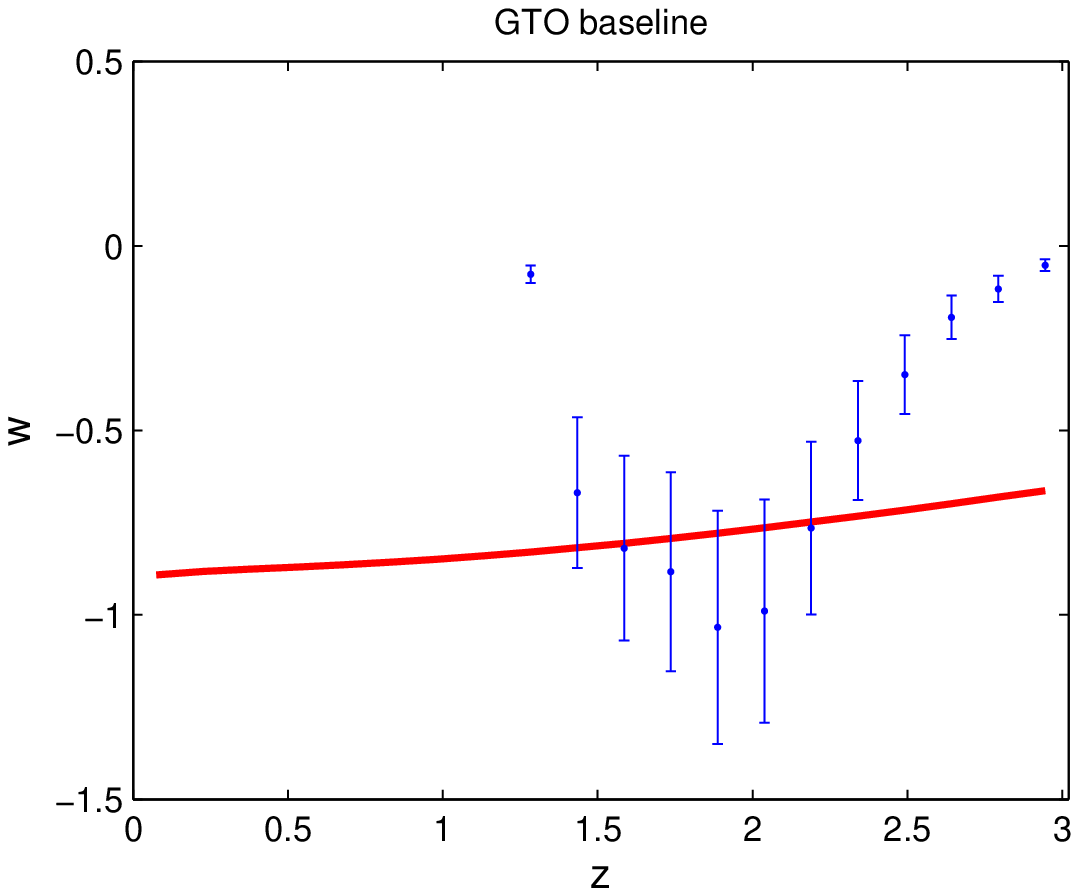}
\includegraphics[width=3.3in]{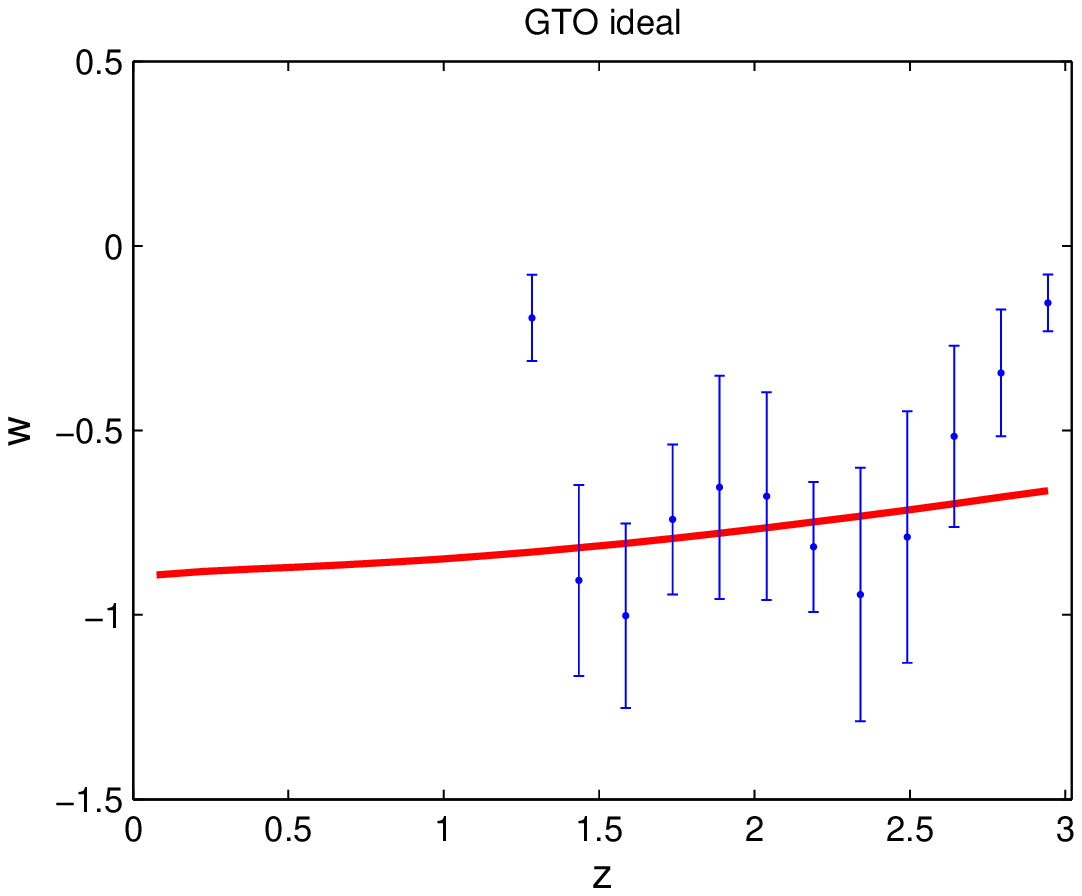}
\includegraphics[width=3.3in]{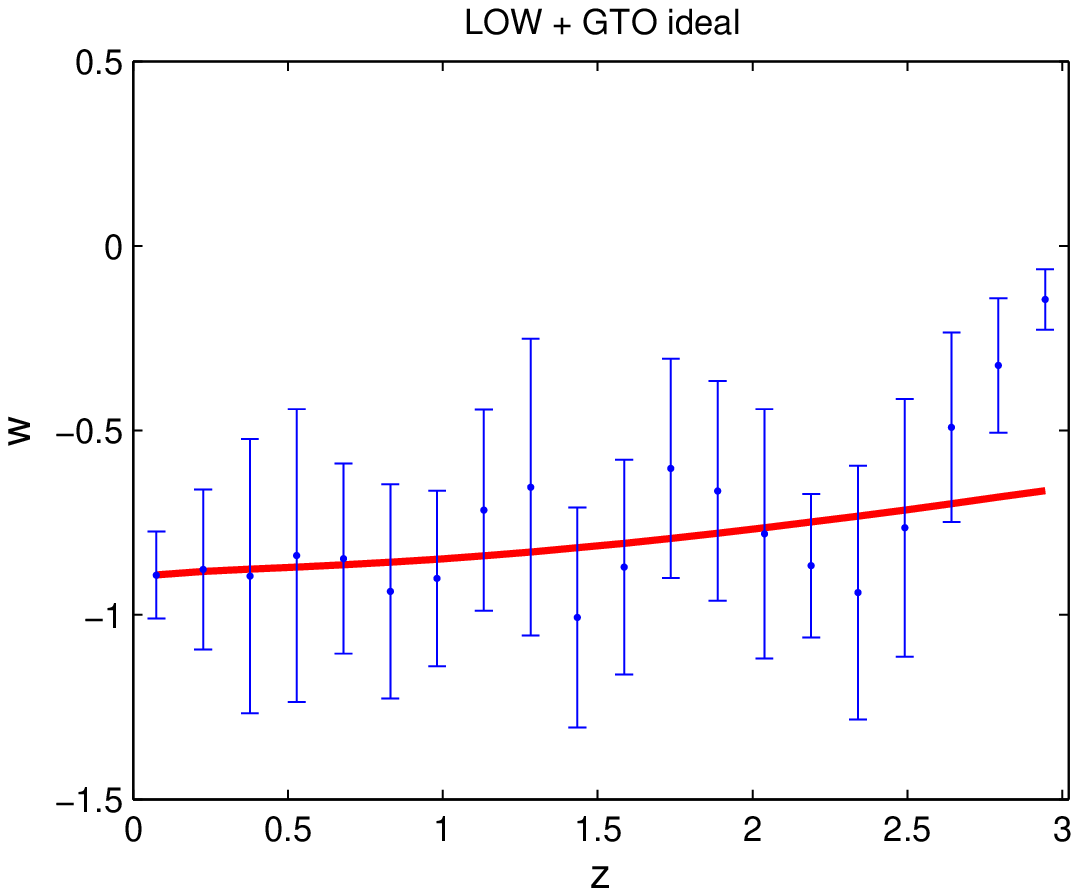}
\includegraphics[width=3.3in]{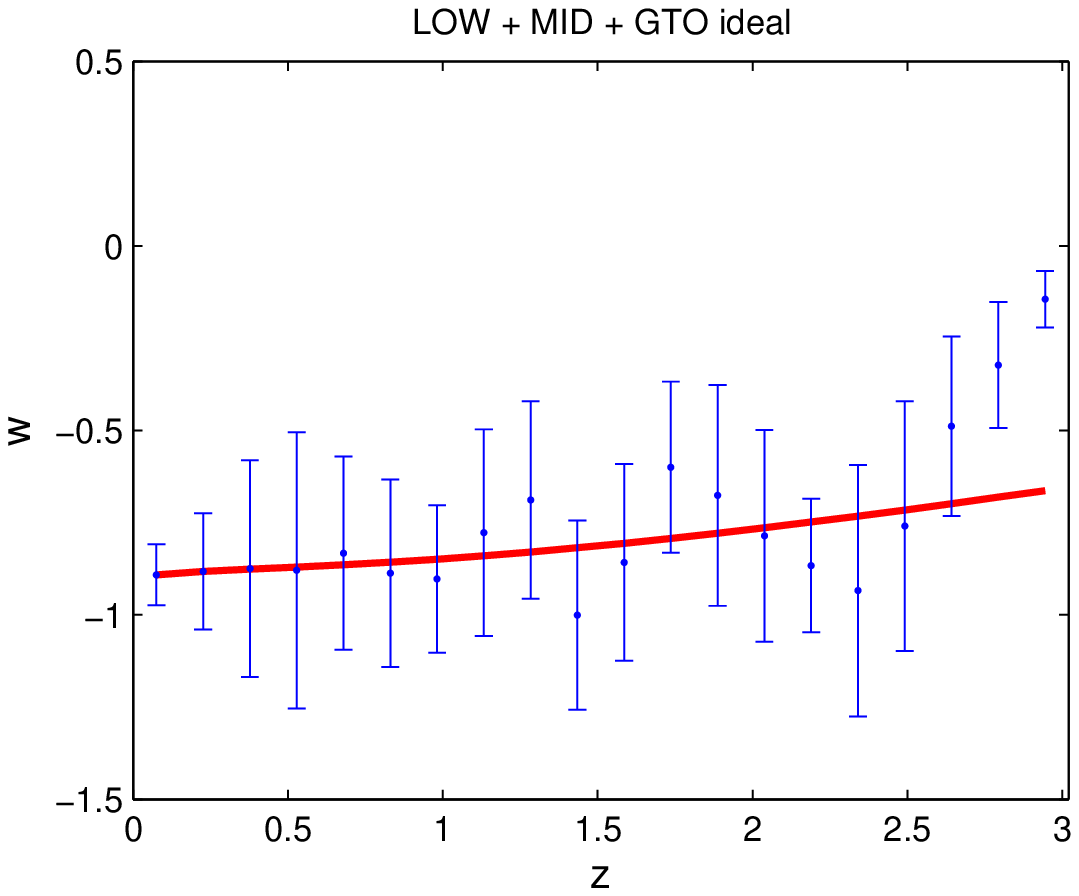}
\end{center}
\caption{Comparing the reconstructed dark energy equation of state $w(z)$ (blue data points) for the EDE equation of state parametrization (red line), for the following cases: LOW supernovas only (top left), LOW plus MID supernovas (top right), ESPRESSO GTO Baseline measurements only (middle left), ESPRESSO GTO ideal data (middle right), LOW supernovas plus ESPRESSO GTO Ideal (bottom left), and LOW plus MID supernovas plus ESPRESSO GTO Ideal (bottom right).}
\label{figwz}
\end{figure*}

Nevertheless, the addition of $\alpha$ measurements to the supernova data does improve the reconstruction, since the extended redshift lever arm enabled by the $\alpha$ measurements improves the determination of the higher modes. This is manifest in the tables from the fact that the ratios ${\cal R}$ are larger for the higher-order modes. Obviously, a post-GTO program extending the dataset can further improve this. Typically the first two PCA modes are not dramatically affected (since these are reasonably well characterized by the low-redshift supernova data), but the higher ones are.

We emphasize that the tables list the gains relative to a supernova-only reconstruction. Since the truncation of the higher modes inherent to the PCA technique will typically produce a biased reconstruction at the highest redshifts probed (e.g., at the highest redshifts the reconstructed equation will approach 0, as was already pointed out in the original work of \cite{HutererandStarkman}), and since this bias will be larger the closer the fiducial equation of state is to a cosmological constant, our $w=-0.9$ case is the one for which the gains are higher. In the CPL and EDE cases, where $w(z)$ increases with redshift, the gains are therefore smaller.

The difference between the baseline and ideal scenarios (which corresponds to a factor of 3 in sensitivity) is noticeable, and this is even more so for the case of ELT-HIRES for which we assumed a gain in sensitivity of a factor of 6. This factor is possibly conservative, though it is difficult to accurately estimate it at this time, given uncertainties such as the E-ELT's transmission in the blue and the wavelength coverage of ELT-HIRES. Still our analysis confirms earlier expectations \cite{Leite1,Leite2} that ELT-HIRES measurements of $\alpha$ can constrain dark energy more strongly than standard supernova surveys.

As a further comparison, Fig. \ref{figmodes} depicts the four best determined eigenmodes for the constant equation of state parametrization, for six of our scenarios, detailed in the figure's caption. All eigenvectors are similarly normalized; how much they differ from zero, in absolute value, indicates their sensitivity to different redshifts (so the change in the overall sign of some modes has no physical significance).

The information gains at high redshift, afforded by the $\alpha$ measurements for the higher modes, are visually clear, corroborating the information in the previously discussed tables. This analysis confirms the well-known result that supernova surveys provide good determinations of two PCA modes (which physically correspond to the present-day dark energy equation of state and its rate of change) but they are ineffective for constraining higher modes. The comparison between the baseline and ideal ESPRESSO datasets (compare the left and right panels in the middle and bottom sets) is also instructive, and confirms that a post-GTO program extending its dataset and improving its sensitivity can have a stronger impact in the field.

Finally, Fig. \ref{figwz} presents examples of the reconstructed dark energy equation of state $w(z)$ (shown by data points with the corresponding error bars) for the EDE equation of state parametrization (always represented by the continuous line), for six of our combinations of simulated datasets detailed in the figure's caption. Note that as discussed in \cite{HutererandStarkman}, the truncation inherent to this reconstruction process is such that the higher redshift bins for which one has data will always lead to an equation of state $w=0$ - so the reconstruction may be biased if the high-redshift behavior of the dark energy equation of state is different. However, as this figure itself makes clear, any such bias can be alleviated by having better data and in particular by extending the redshift coverage.

By comparing the middle and bottom sets of panels in this figure one may naively get the impression that in going from the baseline to the ideal dataset one gets a worst reconstruction. However, this is not the case. The improved sensitivity of the dataset will imply that the optimal number of modes to be used in the dark energy equation of state reconstruction will increase, and while the uncertainties of these PCA higher modes are larger than those of the lower ones (which will correspond to somewhat larger error bars on the reconstructed equation of state), this will also imply that the reconstructed equation of state is much less biased (i.e., much closer to the fiducial one). Therefore in a statistically quantitative sense the reconstruction is indeed s better one.

\section{Conclusion}

A first list of targets for the ESPRESSO fundamental physics GTO has been put together on the basis of the current knowledge of these targets, including the current measurements of $\alpha$ which they lead to. An important complementary task is to assess which among ESPRESSO's modes of operation is optimal for each target. Such a study necessarily requires detailed simulations of ESPRESSO-like spectra, and is currently under way.

The classical way to characterize the redshift dependence of the dark energy equation of state is to use the Hubble diagram from type Ia Supernovas, which at present are limited in redshift: the current maximum is $z\sim1.7$, and measurements at $z>1$ (that is, in the matter era) are currently very scarce. Spectroscopic measurements of $\alpha$ allow us to map dark energy deep in the matter era where, if dark energy is due to a dynamical scalar field, its dynamics is expected to be fastest (and therefore may be easier to detect). Despite ESPRESSO's somewhat limited wavelength coverage as compared to other ESO spectrographs like UVES, it will enable us to characterize dark energy up to a redshift  $z\sim3$.

Previous work using PCA-based forecast techniques  \cite{Amendola1,Leite1,Leite2} had generically shown that the reconstruction of the dark energy equation of state is significantly improved when combining the Type Ia Supernovas with measurements on the stability of the fine-structure constant. Here we have shown that 14 measurements expected from the ESPRESSO fundamental physics GTO will not be able to reconstruct the equation of state and distinguish models in a convincing manner by themselves. However, they will provide important improvements when combined with Type Ia Supernova data. On the other hand, further improvements in sensitivity expected in the Extremely Large Telescopes era will make these measurements a competitive dark energy probe, even on their own. Thus precision astrophysical spectroscopy will soon be a key component of fundamental physics research.

\begin{acknowledgments}
We are grateful to Ana Marta Pinho for helpful discussions on the subject of this work. This work was done in the context of project PTDC/FIS/111725/2009 (FCT, Portugal), with additional support from grant UID/FIS/04434/2013. ACL is supported by an FCT fellowship (SFRH/BD/113746/2015), under the FCT PD Program PhD::SPACE (PD/00040/2012), and by the Gulbenkian Foundation through \textit{Programa de Est\'{i}mulo \`{a} Investiga\c{c}\~{a}o 2014}, grant number 2148613525. CJM is supported by an FCT Research Professorship, contract reference IF/00064/2012, funded by FCT/MCTES (Portugal) and POPH/FSE (EC). SC acknowledges financial contribution from the grant PRIN MIUR 2012 201278X4FL 002 - The Intergalactic Medium as a probe of the growth of cosmic structures.

ACL and CJM thank the Galileo Galilei Institute for Theoretical Physics for the hospitality and the INFN for partial support during the completion of this work.
\end{acknowledgments}

\bibliography{gto}

\begin{thebibliography}{33}%
\makeatletter
\providecommand \@ifxundefined [1]{%
 \@ifx{#1\undefined}
}%
\providecommand \@ifnum [1]{%
 \ifnum #1\expandafter \@firstoftwo
 \else \expandafter \@secondoftwo
 \fi
}%
\providecommand \@ifx [1]{%
 \ifx #1\expandafter \@firstoftwo
 \else \expandafter \@secondoftwo
 \fi
}%
\providecommand \natexlab [1]{#1}%
\providecommand \enquote  [1]{``#1''}%
\providecommand \bibnamefont  [1]{#1}%
\providecommand \bibfnamefont [1]{#1}%
\providecommand \citenamefont [1]{#1}%
\providecommand \href@noop [0]{\@secondoftwo}%
\providecommand \href [0]{\begingroup \@sanitize@url \@href}%
\providecommand \@href[1]{\@@startlink{#1}\@@href}%
\providecommand \@@href[1]{\endgroup#1\@@endlink}%
\providecommand \@sanitize@url [0]{\catcode `\\12\catcode `\$12\catcode
  `\&12\catcode `\#12\catcode `\^12\catcode `\_12\catcode `\%12\relax}%
\providecommand \@@startlink[1]{}%
\providecommand \@@endlink[0]{}%
\providecommand \url  [0]{\begingroup\@sanitize@url \@url }%
\providecommand \@url [1]{\endgroup\@href {#1}{\urlprefix }}%
\providecommand \urlprefix  [0]{URL }%
\providecommand \Eprint [0]{\href }%
\providecommand \doibase [0]{http://dx.doi.org/}%
\providecommand \selectlanguage [0]{\@gobble}%
\providecommand \bibinfo  [0]{\@secondoftwo}%
\providecommand \bibfield  [0]{\@secondoftwo}%
\providecommand \translation [1]{[#1]}%
\providecommand \BibitemOpen [0]{}%
\providecommand \bibitemStop [0]{}%
\providecommand \bibitemNoStop [0]{.\EOS\space}%
\providecommand \EOS [0]{\spacefactor3000\relax}%
\providecommand \BibitemShut  [1]{\csname bibitem#1\endcsname}%
\let\auto@bib@innerbib\@empty
\bibitem [{\citenamefont {Uzan}(2011)}]{Uzan:2010pm}%
  \BibitemOpen
  \bibfield  {author} {\bibinfo {author} {\bibfnamefont {J.-P.}\ \bibnamefont
  {Uzan}},\ }\href@noop {} {\bibfield  {journal} {\bibinfo  {journal} {Living
  Rev.Rel.}\ }\textbf {\bibinfo {volume} {14}},\ \bibinfo {pages} {2} (\bibinfo
  {year} {2011})},\ \Eprint {http://arxiv.org/abs/1009.5514} {arXiv:1009.5514
  [astro-ph.CO]} \BibitemShut {NoStop}%
\bibitem [{\citenamefont {{Pepe}}\ \emph {et~al.}(2013)\citenamefont {{Pepe}},
  \citenamefont {{Cristiani}}, \citenamefont {{Rebolo}}, \citenamefont
  {{Santos}}, \citenamefont {{Dekker}}, \citenamefont {{M{\'e}gevand}},
  \citenamefont {{Zerbi}}, \citenamefont {{Cabral}}, \citenamefont {{Molaro}},
  \citenamefont {{Di Marcantonio}}, \citenamefont {{Abreu}}, \citenamefont
  {{Affolter}}, \citenamefont {{Aliverti}}, \citenamefont {{Allende Prieto}},
  \citenamefont {{Amate}}, \citenamefont {{Avila}}, \citenamefont {{Baldini}},
  \citenamefont {{Bristow}}, \citenamefont {{Broeg}}, \citenamefont {{Cirami}},
  \citenamefont {{Coelho}}, \citenamefont {{Conconi}}, \citenamefont
  {{Coretti}}, \citenamefont {{Cupani}}, \citenamefont {{D'Odorico}},
  \citenamefont {{De Caprio}}, \citenamefont {{Delabre}}, \citenamefont
  {{Dorn}}, \citenamefont {{Figueira}}, \citenamefont {{Fragoso}},
  \citenamefont {{Galeotta}}, \citenamefont {{Genolet}}, \citenamefont
  {{Gomes}}, \citenamefont {{Gonz{\'a}lez Hern{\'a}ndez}}, \citenamefont
  {{Hughes}}, \citenamefont {{Iwert}}, \citenamefont {{Kerber}}, \citenamefont
  {{Landoni}}, \citenamefont {{Lizon}}, \citenamefont {{Lovis}}, \citenamefont
  {{Maire}}, \citenamefont {{Mannetta}}, \citenamefont {{Martins}},
  \citenamefont {{Monteiro}}, \citenamefont {{Oliveira}}, \citenamefont
  {{Poretti}}, \citenamefont {{Rasilla}}, \citenamefont {{Riva}}, \citenamefont
  {{Santana Tschudi}}, \citenamefont {{Santos}}, \citenamefont {{Sosnowska}},
  \citenamefont {{Sousa}}, \citenamefont {{Span{\`o}}}, \citenamefont
  {{Tenegi}}, \citenamefont {{Toso}}, \citenamefont {{Vanzella}}, \citenamefont
  {{Viel}},\ and\ \citenamefont {{Zapatero Osorio}}}]{espresso}%
  \BibitemOpen
  \bibfield  {author} {\bibinfo {author} {\bibfnamefont {F.}~\bibnamefont
  {{Pepe}}}, \bibinfo {author} {\bibfnamefont {S.}~\bibnamefont {{Cristiani}}},
  \bibinfo {author} {\bibfnamefont {R.}~\bibnamefont {{Rebolo}}}, \bibinfo
  {author} {\bibfnamefont {N.~C.}\ \bibnamefont {{Santos}}}, \bibinfo {author}
  {\bibfnamefont {H.}~\bibnamefont {{Dekker}}}, \bibinfo {author}
  {\bibfnamefont {D.}~\bibnamefont {{M{\'e}gevand}}}, \bibinfo {author}
  {\bibfnamefont {F.~M.}\ \bibnamefont {{Zerbi}}}, \bibinfo {author}
  {\bibfnamefont {A.}~\bibnamefont {{Cabral}}}, \bibinfo {author}
  {\bibfnamefont {P.}~\bibnamefont {{Molaro}}}, \bibinfo {author}
  {\bibfnamefont {P.}~\bibnamefont {{Di Marcantonio}}}, \bibinfo {author}
  {\bibfnamefont {M.}~\bibnamefont {{Abreu}}}, \bibinfo {author} {\bibfnamefont
  {M.}~\bibnamefont {{Affolter}}}, \bibinfo {author} {\bibfnamefont
  {M.}~\bibnamefont {{Aliverti}}}, \bibinfo {author} {\bibfnamefont
  {C.}~\bibnamefont {{Allende Prieto}}}, \bibinfo {author} {\bibfnamefont
  {M.}~\bibnamefont {{Amate}}}, \bibinfo {author} {\bibfnamefont
  {G.}~\bibnamefont {{Avila}}}, \bibinfo {author} {\bibfnamefont
  {V.}~\bibnamefont {{Baldini}}}, \bibinfo {author} {\bibfnamefont
  {P.}~\bibnamefont {{Bristow}}}, \bibinfo {author} {\bibfnamefont
  {C.}~\bibnamefont {{Broeg}}}, \bibinfo {author} {\bibfnamefont
  {R.}~\bibnamefont {{Cirami}}}, \bibinfo {author} {\bibfnamefont
  {J.}~\bibnamefont {{Coelho}}}, \bibinfo {author} {\bibfnamefont
  {P.}~\bibnamefont {{Conconi}}}, \bibinfo {author} {\bibfnamefont
  {I.}~\bibnamefont {{Coretti}}}, \bibinfo {author} {\bibfnamefont
  {G.}~\bibnamefont {{Cupani}}}, \bibinfo {author} {\bibfnamefont
  {V.}~\bibnamefont {{D'Odorico}}}, \bibinfo {author} {\bibfnamefont
  {V.}~\bibnamefont {{De Caprio}}}, \bibinfo {author} {\bibfnamefont
  {B.}~\bibnamefont {{Delabre}}}, \bibinfo {author} {\bibfnamefont
  {R.}~\bibnamefont {{Dorn}}}, \bibinfo {author} {\bibfnamefont
  {P.}~\bibnamefont {{Figueira}}}, \bibinfo {author} {\bibfnamefont
  {A.}~\bibnamefont {{Fragoso}}}, \bibinfo {author} {\bibfnamefont
  {S.}~\bibnamefont {{Galeotta}}}, \bibinfo {author} {\bibfnamefont
  {L.}~\bibnamefont {{Genolet}}}, \bibinfo {author} {\bibfnamefont
  {R.}~\bibnamefont {{Gomes}}}, \bibinfo {author} {\bibfnamefont {J.~I.}\
  \bibnamefont {{Gonz{\'a}lez Hern{\'a}ndez}}}, \bibinfo {author}
  {\bibfnamefont {I.}~\bibnamefont {{Hughes}}}, \bibinfo {author}
  {\bibfnamefont {O.}~\bibnamefont {{Iwert}}}, \bibinfo {author} {\bibfnamefont
  {F.}~\bibnamefont {{Kerber}}}, \bibinfo {author} {\bibfnamefont
  {M.}~\bibnamefont {{Landoni}}}, \bibinfo {author} {\bibfnamefont {J.-L.}\
  \bibnamefont {{Lizon}}}, \bibinfo {author} {\bibfnamefont {C.}~\bibnamefont
  {{Lovis}}}, \bibinfo {author} {\bibfnamefont {C.}~\bibnamefont {{Maire}}},
  \bibinfo {author} {\bibfnamefont {M.}~\bibnamefont {{Mannetta}}}, \bibinfo
  {author} {\bibfnamefont {C.}~\bibnamefont {{Martins}}}, \bibinfo {author}
  {\bibfnamefont {M.~A.}\ \bibnamefont {{Monteiro}}}, \bibinfo {author}
  {\bibfnamefont {A.}~\bibnamefont {{Oliveira}}}, \bibinfo {author}
  {\bibfnamefont {E.}~\bibnamefont {{Poretti}}}, \bibinfo {author}
  {\bibfnamefont {J.~L.}\ \bibnamefont {{Rasilla}}}, \bibinfo {author}
  {\bibfnamefont {M.}~\bibnamefont {{Riva}}}, \bibinfo {author} {\bibfnamefont
  {S.}~\bibnamefont {{Santana Tschudi}}}, \bibinfo {author} {\bibfnamefont
  {P.}~\bibnamefont {{Santos}}}, \bibinfo {author} {\bibfnamefont
  {D.}~\bibnamefont {{Sosnowska}}}, \bibinfo {author} {\bibfnamefont
  {S.}~\bibnamefont {{Sousa}}}, \bibinfo {author} {\bibfnamefont
  {P.}~\bibnamefont {{Span{\`o}}}}, \bibinfo {author} {\bibfnamefont
  {F.}~\bibnamefont {{Tenegi}}}, \bibinfo {author} {\bibfnamefont
  {G.}~\bibnamefont {{Toso}}}, \bibinfo {author} {\bibfnamefont
  {E.}~\bibnamefont {{Vanzella}}}, \bibinfo {author} {\bibfnamefont
  {M.}~\bibnamefont {{Viel}}}, \ and\ \bibinfo {author} {\bibfnamefont {M.~R.}\
  \bibnamefont {{Zapatero Osorio}}},\ }\href@noop {} {\bibfield  {journal}
  {\bibinfo  {journal} {The Messenger}\ }\textbf {\bibinfo {volume} {153}},\
  \bibinfo {pages} {6} (\bibinfo {year} {2013})}\BibitemShut {NoStop}%
\bibitem [{\citenamefont {Webb}\ \emph {et~al.}(2011)\citenamefont {Webb},
  \citenamefont {King}, \citenamefont {Murphy}, \citenamefont {Flambaum},
  \citenamefont {Carswell},\ and\ \citenamefont {Bainbridge}}]{Dipole}%
  \BibitemOpen
  \bibfield  {author} {\bibinfo {author} {\bibfnamefont {J.~K.}\ \bibnamefont
  {Webb}}, \bibinfo {author} {\bibfnamefont {J.~A.}\ \bibnamefont {King}},
  \bibinfo {author} {\bibfnamefont {M.~T.}\ \bibnamefont {Murphy}}, \bibinfo
  {author} {\bibfnamefont {V.~V.}\ \bibnamefont {Flambaum}}, \bibinfo {author}
  {\bibfnamefont {R.~F.}\ \bibnamefont {Carswell}}, \ and\ \bibinfo {author}
  {\bibfnamefont {M.~B.}\ \bibnamefont {Bainbridge}},\ }\href {\doibase
  10.1103/PhysRevLett.107.191101} {\bibfield  {journal} {\bibinfo  {journal}
  {Phys. Rev. Lett.}\ }\textbf {\bibinfo {volume} {107}},\ \bibinfo {pages}
  {191101} (\bibinfo {year} {2011})},\ \Eprint {http://arxiv.org/abs/1008.3907}
  {arXiv:1008.3907 [astro-ph.CO]} \BibitemShut {NoStop}%
\bibitem [{\citenamefont {Martins}(2015)}]{GRG}%
  \BibitemOpen
  \bibfield  {author} {\bibinfo {author} {\bibfnamefont {C.~J. A.~P.}\
  \bibnamefont {Martins}},\ }\href {\doibase 10.1007/s10714-014-1843-7}
  {\bibfield  {journal} {\bibinfo  {journal} {Gen. Rel. Grav.}\ }\textbf
  {\bibinfo {volume} {47}},\ \bibinfo {pages} {1843} (\bibinfo {year}
  {2015})},\ \Eprint {http://arxiv.org/abs/1412.0108} {arXiv:1412.0108
  [astro-ph.CO]} \BibitemShut {NoStop}%
\bibitem [{\citenamefont {{Amendola}}\ \emph {et~al.}(2012)\citenamefont
  {{Amendola}}, \citenamefont {{Leite}}, \citenamefont {{Martins}},
  \citenamefont {{Nunes}}, \citenamefont {{Pedrosa}},\ and\ \citenamefont
  {{Seganti}}}]{Amendola1}%
  \BibitemOpen
  \bibfield  {author} {\bibinfo {author} {\bibfnamefont {L.}~\bibnamefont
  {{Amendola}}}, \bibinfo {author} {\bibfnamefont {A.~C.~O.}\ \bibnamefont
  {{Leite}}}, \bibinfo {author} {\bibfnamefont {C.~J.~A.~P.}\ \bibnamefont
  {{Martins}}}, \bibinfo {author} {\bibfnamefont {N.~J.}\ \bibnamefont
  {{Nunes}}}, \bibinfo {author} {\bibfnamefont {P.~O.~J.}\ \bibnamefont
  {{Pedrosa}}}, \ and\ \bibinfo {author} {\bibfnamefont {A.}~\bibnamefont
  {{Seganti}}},\ }\href {\doibase 10.1103/PhysRevD.86.063515} {\bibfield
  {journal} {\bibinfo  {journal} {\prd}\ }\textbf {\bibinfo {volume} {86}},\
  \bibinfo {eid} {063515} (\bibinfo {year} {2012})},\ \Eprint
  {http://arxiv.org/abs/1109.6793} {arXiv:1109.6793} \BibitemShut {NoStop}%
\bibitem [{\citenamefont {{Leite}}\ \emph {et~al.}(2014)\citenamefont
  {{Leite}}, \citenamefont {{Martins}}, \citenamefont {{Pedrosa}},\ and\
  \citenamefont {{Nunes}}}]{Leite1}%
  \BibitemOpen
  \bibfield  {author} {\bibinfo {author} {\bibfnamefont {A.~C.~O.}\
  \bibnamefont {{Leite}}}, \bibinfo {author} {\bibfnamefont {C.~J.~A.~P.}\
  \bibnamefont {{Martins}}}, \bibinfo {author} {\bibfnamefont {P.~O.~J.}\
  \bibnamefont {{Pedrosa}}}, \ and\ \bibinfo {author} {\bibfnamefont {N.~J.}\
  \bibnamefont {{Nunes}}},\ }\href {\doibase 10.1103/PhysRevD.90.063519}
  {\bibfield  {journal} {\bibinfo  {journal} {\prd}\ }\textbf {\bibinfo
  {volume} {90}},\ \bibinfo {eid} {063519} (\bibinfo {year} {2014})},\ \Eprint
  {http://arxiv.org/abs/1409.3963} {arXiv:1409.3963} \BibitemShut {NoStop}%
\bibitem [{\citenamefont {{Leite}}\ and\ \citenamefont
  {{Martins}}(2015)}]{Leite2}%
  \BibitemOpen
  \bibfield  {author} {\bibinfo {author} {\bibfnamefont {A.~C.~O.}\
  \bibnamefont {{Leite}}}\ and\ \bibinfo {author} {\bibfnamefont {C.~J.~A.~P.}\
  \bibnamefont {{Martins}}},\ }\href {\doibase 10.1103/PhysRevD.91.103519}
  {\bibfield  {journal} {\bibinfo  {journal} {\prd}\ }\textbf {\bibinfo
  {volume} {91}},\ \bibinfo {eid} {103519} (\bibinfo {year} {2015})},\ \Eprint
  {http://arxiv.org/abs/1505.05529} {arXiv:1505.05529} \BibitemShut {NoStop}%
\bibitem [{\citenamefont {Martins}\ \emph {et~al.}(2016)\citenamefont
  {Martins}, \citenamefont {Pinho}, \citenamefont {Carreira}, \citenamefont
  {Gusart}, \citenamefont {L\'opez},\ and\ \citenamefont {Rocha}}]{Pinho}%
  \BibitemOpen
  \bibfield  {author} {\bibinfo {author} {\bibfnamefont {C.~J. A.~P.}\
  \bibnamefont {Martins}}, \bibinfo {author} {\bibfnamefont {A.~M.~M.}\
  \bibnamefont {Pinho}}, \bibinfo {author} {\bibfnamefont {P.}~\bibnamefont
  {Carreira}}, \bibinfo {author} {\bibfnamefont {A.}~\bibnamefont {Gusart}},
  \bibinfo {author} {\bibfnamefont {J.}~\bibnamefont {L\'opez}}, \ and\
  \bibinfo {author} {\bibfnamefont {C.~I. S.~A.}\ \bibnamefont {Rocha}},\
  }\href {\doibase 10.1103/PhysRevD.93.023506} {\bibfield  {journal} {\bibinfo
  {journal} {Phys. Rev.}\ }\textbf {\bibinfo {volume} {D93}},\ \bibinfo {pages}
  {023506} (\bibinfo {year} {2016})},\ \Eprint
  {http://arxiv.org/abs/1601.02950} {arXiv:1601.02950 [astro-ph.CO]}
  \BibitemShut {NoStop}%
\bibitem [{\citenamefont {{Huterer}}\ and\ \citenamefont
  {{Starkman}}(2003)}]{HutererandStarkman}%
  \BibitemOpen
  \bibfield  {author} {\bibinfo {author} {\bibfnamefont {D.}~\bibnamefont
  {{Huterer}}}\ and\ \bibinfo {author} {\bibfnamefont {G.}~\bibnamefont
  {{Starkman}}},\ }\href {\doibase 10.1103/PhysRevLett.90.031301} {\bibfield
  {journal} {\bibinfo  {journal} {Physical Review Letters}\ }\textbf {\bibinfo
  {volume} {90}},\ \bibinfo {eid} {031301} (\bibinfo {year} {2003})},\ \Eprint
  {http://arxiv.org/abs/astro-ph/0207517} {astro-ph/0207517} \BibitemShut
  {NoStop}%
\bibitem [{\citenamefont {{Albrecht}}\ \emph {et~al.}(2009)\citenamefont
  {{Albrecht}}, \citenamefont {{Amendola}}, \citenamefont {{Bernstein}},
  \citenamefont {{Clowe}}, \citenamefont {{Eisenstein}}, \citenamefont
  {{Guzzo}}, \citenamefont {{Hirata}}, \citenamefont {{Huterer}}, \citenamefont
  {{Kirshner}}, \citenamefont {{Kolb}},\ and\ \citenamefont {{Nichol}}}]{DEFM}%
  \BibitemOpen
  \bibfield  {author} {\bibinfo {author} {\bibfnamefont {A.}~\bibnamefont
  {{Albrecht}}}, \bibinfo {author} {\bibfnamefont {L.}~\bibnamefont
  {{Amendola}}}, \bibinfo {author} {\bibfnamefont {G.}~\bibnamefont
  {{Bernstein}}}, \bibinfo {author} {\bibfnamefont {D.}~\bibnamefont
  {{Clowe}}}, \bibinfo {author} {\bibfnamefont {D.}~\bibnamefont
  {{Eisenstein}}}, \bibinfo {author} {\bibfnamefont {L.}~\bibnamefont
  {{Guzzo}}}, \bibinfo {author} {\bibfnamefont {C.}~\bibnamefont {{Hirata}}},
  \bibinfo {author} {\bibfnamefont {D.}~\bibnamefont {{Huterer}}}, \bibinfo
  {author} {\bibfnamefont {R.}~\bibnamefont {{Kirshner}}}, \bibinfo {author}
  {\bibfnamefont {E.}~\bibnamefont {{Kolb}}}, \ and\ \bibinfo {author}
  {\bibfnamefont {R.}~\bibnamefont {{Nichol}}},\ }\href@noop {} {\bibfield
  {journal} {\bibinfo  {journal} {ArXiv e-prints}\ } (\bibinfo {year}
  {2009})},\ \Eprint {http://arxiv.org/abs/0901.0721} {arXiv:0901.0721
  [astro-ph.IM]} \BibitemShut {NoStop}%
\bibitem [{\citenamefont {Liske}\ \emph {et~al.}(2014)\citenamefont {Liske}
  \emph {et~al.}}]{HIRES}%
  \BibitemOpen
  \bibfield  {author} {\bibinfo {author} {\bibfnamefont {J.}~\bibnamefont
  {Liske}} \emph {et~al.},\ }\href@noop {} {\emph {\bibinfo {title} {{Top Level
  Requirements For ELT-HIRES}}}},\ \bibinfo {type} {Tech. Rep.}\ (\bibinfo
  {institution} {Document ESO 204697 Version 1},\ \bibinfo {year}
  {2014})\BibitemShut {NoStop}%
\bibitem [{\citenamefont {Dzuba}\ \emph
  {et~al.}(1999{\natexlab{a}})\citenamefont {Dzuba}, \citenamefont {Flambaum},\
  and\ \citenamefont {Webb}}]{Dzuba1}%
  \BibitemOpen
  \bibfield  {author} {\bibinfo {author} {\bibfnamefont {V.~A.}\ \bibnamefont
  {Dzuba}}, \bibinfo {author} {\bibfnamefont {V.~V.}\ \bibnamefont {Flambaum}},
  \ and\ \bibinfo {author} {\bibfnamefont {J.~K.}\ \bibnamefont {Webb}},\
  }\href {\doibase 10.1103/PhysRevLett.82.888} {\bibfield  {journal} {\bibinfo
  {journal} {Phys. Rev. Lett.}\ }\textbf {\bibinfo {volume} {82}},\ \bibinfo
  {pages} {888} (\bibinfo {year} {1999}{\natexlab{a}})},\ \Eprint
  {http://arxiv.org/abs/physics/9802029} {arXiv:physics/9802029
  [physics.atom-ph]} \BibitemShut {NoStop}%
\bibitem [{\citenamefont {Dzuba}\ \emph
  {et~al.}(1999{\natexlab{b}})\citenamefont {Dzuba}, \citenamefont {Flambaum},\
  and\ \citenamefont {Webb}}]{Dzuba2}%
  \BibitemOpen
  \bibfield  {author} {\bibinfo {author} {\bibfnamefont {V.~A.}\ \bibnamefont
  {Dzuba}}, \bibinfo {author} {\bibfnamefont {V.~V.}\ \bibnamefont {Flambaum}},
  \ and\ \bibinfo {author} {\bibfnamefont {J.~K.}\ \bibnamefont {Webb}},\
  }\href {\doibase 10.1103/PhysRevA.59.230} {\bibfield  {journal} {\bibinfo
  {journal} {Phys. Rev.}\ }\textbf {\bibinfo {volume} {A59}},\ \bibinfo {pages}
  {230} (\bibinfo {year} {1999}{\natexlab{b}})},\ \Eprint
  {http://arxiv.org/abs/physics/9808021} {arXiv:physics/9808021 [physics]}
  \BibitemShut {NoStop}%
\bibitem [{\citenamefont {{King}}(2012)}]{king}%
  \BibitemOpen
  \bibfield  {author} {\bibinfo {author} {\bibfnamefont {J.~A.}\ \bibnamefont
  {{King}}},\ }\emph {\bibinfo {title} {{Searching for variations in the
  fine-structure constant and the proton-to-electron mass ratio using quasar
  absorption lines}}},\ \href@noop {} {Ph.D. thesis},\ \bibinfo  {school}
  {School of Physics, The University of New South Wales} (\bibinfo {year}
  {2012})\BibitemShut {NoStop}%
\bibitem [{\citenamefont {{Murphy}}(2002)}]{murphy}%
  \BibitemOpen
  \bibfield  {author} {\bibinfo {author} {\bibfnamefont {M.~T.}\ \bibnamefont
  {{Murphy}}},\ }\emph {\bibinfo {title} {{Probing variations in the
  fundamental constants with quasar absorption lines}}},\ \href@noop {} {Ph.D.
  thesis},\ \bibinfo  {school} {Institute of Astronomy, University of
  Cambridge} (\bibinfo {year} {2002})\BibitemShut {NoStop}%
\bibitem [{\citenamefont {{Chand}}\ \emph {et~al.}(2004)\citenamefont
  {{Chand}}, \citenamefont {{Srianand}}, \citenamefont {{Petitjean}},\ and\
  \citenamefont {{Aracil}}}]{chand1}%
  \BibitemOpen
  \bibfield  {author} {\bibinfo {author} {\bibfnamefont {H.}~\bibnamefont
  {{Chand}}}, \bibinfo {author} {\bibfnamefont {R.}~\bibnamefont {{Srianand}}},
  \bibinfo {author} {\bibfnamefont {P.}~\bibnamefont {{Petitjean}}}, \ and\
  \bibinfo {author} {\bibfnamefont {B.}~\bibnamefont {{Aracil}}},\ }\href
  {\doibase 10.1051/0004-6361:20035701} {\bibfield  {journal} {\bibinfo
  {journal} {A. \& A.}\ }\textbf {\bibinfo {volume} {417}},\ \bibinfo {pages}
  {853} (\bibinfo {year} {2004})},\ \Eprint
  {http://arxiv.org/abs/astro-ph/0401094} {astro-ph/0401094} \BibitemShut
  {NoStop}%
\bibitem [{\citenamefont {{Chand}}\ \emph {et~al.}(2005)\citenamefont
  {{Chand}}, \citenamefont {{Petitjean}}, \citenamefont {{Srianand}},\ and\
  \citenamefont {{Aracil}}}]{chand2}%
  \BibitemOpen
  \bibfield  {author} {\bibinfo {author} {\bibfnamefont {H.}~\bibnamefont
  {{Chand}}}, \bibinfo {author} {\bibfnamefont {P.}~\bibnamefont
  {{Petitjean}}}, \bibinfo {author} {\bibfnamefont {R.}~\bibnamefont
  {{Srianand}}}, \ and\ \bibinfo {author} {\bibfnamefont {B.}~\bibnamefont
  {{Aracil}}},\ }\href {\doibase 10.1051/0004-6361:20041186} {\bibfield
  {journal} {\bibinfo  {journal} {A. \& A.}\ }\textbf {\bibinfo {volume}
  {430}},\ \bibinfo {pages} {47} (\bibinfo {year} {2005})},\ \Eprint
  {http://arxiv.org/abs/astro-ph/0408200} {astro-ph/0408200} \BibitemShut
  {NoStop}%
\bibitem [{\citenamefont {{Chand}}\ \emph {et~al.}(2006)\citenamefont
  {{Chand}}, \citenamefont {{Srianand}}, \citenamefont {{Petitjean}},
  \citenamefont {{Aracil}}, \citenamefont {{Quast}},\ and\ \citenamefont
  {{Reimers}}}]{chand3}%
  \BibitemOpen
  \bibfield  {author} {\bibinfo {author} {\bibfnamefont {H.}~\bibnamefont
  {{Chand}}}, \bibinfo {author} {\bibfnamefont {R.}~\bibnamefont {{Srianand}}},
  \bibinfo {author} {\bibfnamefont {P.}~\bibnamefont {{Petitjean}}}, \bibinfo
  {author} {\bibfnamefont {B.}~\bibnamefont {{Aracil}}}, \bibinfo {author}
  {\bibfnamefont {R.}~\bibnamefont {{Quast}}}, \ and\ \bibinfo {author}
  {\bibfnamefont {D.}~\bibnamefont {{Reimers}}},\ }\href {\doibase
  10.1051/0004-6361:20054584} {\bibfield  {journal} {\bibinfo  {journal} {A. \&
  A.}\ }\textbf {\bibinfo {volume} {451}},\ \bibinfo {pages} {45} (\bibinfo
  {year} {2006})},\ \Eprint {http://arxiv.org/abs/astro-ph/0601194}
  {astro-ph/0601194} \BibitemShut {NoStop}%
\bibitem [{\citenamefont {{Levshakov}}\ \emph {et~al.}(2007)\citenamefont
  {{Levshakov}}, \citenamefont {{Molaro}}, \citenamefont {{Lopez}},
  \citenamefont {{D'Odorico}}, \citenamefont {{Centuri{\'o}n}}, \citenamefont
  {{Bonifacio}}, \citenamefont {{Agafonova}},\ and\ \citenamefont
  {{Reimers}}}]{lev1}%
  \BibitemOpen
  \bibfield  {author} {\bibinfo {author} {\bibfnamefont {S.~A.}\ \bibnamefont
  {{Levshakov}}}, \bibinfo {author} {\bibfnamefont {P.}~\bibnamefont
  {{Molaro}}}, \bibinfo {author} {\bibfnamefont {S.}~\bibnamefont {{Lopez}}},
  \bibinfo {author} {\bibfnamefont {S.}~\bibnamefont {{D'Odorico}}}, \bibinfo
  {author} {\bibfnamefont {M.}~\bibnamefont {{Centuri{\'o}n}}}, \bibinfo
  {author} {\bibfnamefont {P.}~\bibnamefont {{Bonifacio}}}, \bibinfo {author}
  {\bibfnamefont {I.~I.}\ \bibnamefont {{Agafonova}}}, \ and\ \bibinfo {author}
  {\bibfnamefont {D.}~\bibnamefont {{Reimers}}},\ }\href {\doibase
  10.1051/0004-6361:20066064} {\bibfield  {journal} {\bibinfo  {journal} {A. \&
  A.}\ }\textbf {\bibinfo {volume} {466}},\ \bibinfo {pages} {1077} (\bibinfo
  {year} {2007})},\ \Eprint {http://arxiv.org/abs/astro-ph/0703042}
  {astro-ph/0703042} \BibitemShut {NoStop}%
\bibitem [{\citenamefont {{Molaro}}\ \emph {et~al.}(2008)\citenamefont
  {{Molaro}}, \citenamefont {{Reimers}}, \citenamefont {{Agafonova}},\ and\
  \citenamefont {{Levshakov}}}]{lev2}%
  \BibitemOpen
  \bibfield  {author} {\bibinfo {author} {\bibfnamefont {P.}~\bibnamefont
  {{Molaro}}}, \bibinfo {author} {\bibfnamefont {D.}~\bibnamefont {{Reimers}}},
  \bibinfo {author} {\bibfnamefont {I.~I.}\ \bibnamefont {{Agafonova}}}, \ and\
  \bibinfo {author} {\bibfnamefont {S.~A.}\ \bibnamefont {{Levshakov}}},\
  }\href {\doibase 10.1140/epjst/e2008-00818-4} {\bibfield  {journal} {\bibinfo
   {journal} {European Physical Journal Special Topics}\ }\textbf {\bibinfo
  {volume} {163}},\ \bibinfo {pages} {173} (\bibinfo {year} {2008})},\ \Eprint
  {http://arxiv.org/abs/0712.4380} {arXiv:0712.4380} \BibitemShut {NoStop}%
\bibitem [{\citenamefont {{Molaro}}\ \emph {et~al.}(2013)\citenamefont
  {{Molaro}}, \citenamefont {{Centuri{\'o}n}}, \citenamefont {{Whitmore}},
  \citenamefont {{Evans}}, \citenamefont {{Murphy}}, \citenamefont
  {{Agafonova}}, \citenamefont {{Bonifacio}}, \citenamefont {{D'Odorico}},
  \citenamefont {{Levshakov}}, \citenamefont {{Lopez}}, \citenamefont
  {{Martins}}, \citenamefont {{Petitjean}}, \citenamefont {{Rahmani}},
  \citenamefont {{Reimers}}, \citenamefont {{Srianand}}, \citenamefont
  {{Vladilo}},\ and\ \citenamefont {{Wendt}}}]{lp1}%
  \BibitemOpen
  \bibfield  {author} {\bibinfo {author} {\bibfnamefont {P.}~\bibnamefont
  {{Molaro}}}, \bibinfo {author} {\bibfnamefont {M.}~\bibnamefont
  {{Centuri{\'o}n}}}, \bibinfo {author} {\bibfnamefont {J.~B.}\ \bibnamefont
  {{Whitmore}}}, \bibinfo {author} {\bibfnamefont {T.~M.}\ \bibnamefont
  {{Evans}}}, \bibinfo {author} {\bibfnamefont {M.~T.}\ \bibnamefont
  {{Murphy}}}, \bibinfo {author} {\bibfnamefont {I.~I.}\ \bibnamefont
  {{Agafonova}}}, \bibinfo {author} {\bibfnamefont {P.}~\bibnamefont
  {{Bonifacio}}}, \bibinfo {author} {\bibfnamefont {S.}~\bibnamefont
  {{D'Odorico}}}, \bibinfo {author} {\bibfnamefont {S.~A.}\ \bibnamefont
  {{Levshakov}}}, \bibinfo {author} {\bibfnamefont {S.}~\bibnamefont
  {{Lopez}}}, \bibinfo {author} {\bibfnamefont {C.~J.~A.~P.}\ \bibnamefont
  {{Martins}}}, \bibinfo {author} {\bibfnamefont {P.}~\bibnamefont
  {{Petitjean}}}, \bibinfo {author} {\bibfnamefont {H.}~\bibnamefont
  {{Rahmani}}}, \bibinfo {author} {\bibfnamefont {D.}~\bibnamefont
  {{Reimers}}}, \bibinfo {author} {\bibfnamefont {R.}~\bibnamefont
  {{Srianand}}}, \bibinfo {author} {\bibfnamefont {G.}~\bibnamefont
  {{Vladilo}}}, \ and\ \bibinfo {author} {\bibfnamefont {M.}~\bibnamefont
  {{Wendt}}},\ }\href {\doibase 10.1051/0004-6361/201321351} {\bibfield
  {journal} {\bibinfo  {journal} {A. \& A.}\ }\textbf {\bibinfo {volume}
  {555}},\ \bibinfo {eid} {A68} (\bibinfo {year} {2013})},\ \Eprint
  {http://arxiv.org/abs/1305.1884} {arXiv:1305.1884} \BibitemShut {NoStop}%
\bibitem [{\citenamefont {{Rahmani}}\ \emph {et~al.}(2013)\citenamefont
  {{Rahmani}}, \citenamefont {{Wendt}}, \citenamefont {{Srianand}},
  \citenamefont {{Noterdaeme}}, \citenamefont {{Petitjean}}, \citenamefont
  {{Molaro}}, \citenamefont {{Whitmore}}, \citenamefont {{Murphy}},
  \citenamefont {{Centurion}}, \citenamefont {{Fathivavsari}}, \citenamefont
  {{D'Odorico}}, \citenamefont {{Evans}}, \citenamefont {{Levshakov}},
  \citenamefont {{Lopez}}, \citenamefont {{Martins}}, \citenamefont
  {{Reimers}},\ and\ \citenamefont {{Vladilo}}}]{lp2}%
  \BibitemOpen
  \bibfield  {author} {\bibinfo {author} {\bibfnamefont {H.}~\bibnamefont
  {{Rahmani}}}, \bibinfo {author} {\bibfnamefont {M.}~\bibnamefont {{Wendt}}},
  \bibinfo {author} {\bibfnamefont {R.}~\bibnamefont {{Srianand}}}, \bibinfo
  {author} {\bibfnamefont {P.}~\bibnamefont {{Noterdaeme}}}, \bibinfo {author}
  {\bibfnamefont {P.}~\bibnamefont {{Petitjean}}}, \bibinfo {author}
  {\bibfnamefont {P.}~\bibnamefont {{Molaro}}}, \bibinfo {author}
  {\bibfnamefont {J.~B.}\ \bibnamefont {{Whitmore}}}, \bibinfo {author}
  {\bibfnamefont {M.~T.}\ \bibnamefont {{Murphy}}}, \bibinfo {author}
  {\bibfnamefont {M.}~\bibnamefont {{Centurion}}}, \bibinfo {author}
  {\bibfnamefont {H.}~\bibnamefont {{Fathivavsari}}}, \bibinfo {author}
  {\bibfnamefont {S.}~\bibnamefont {{D'Odorico}}}, \bibinfo {author}
  {\bibfnamefont {T.~M.}\ \bibnamefont {{Evans}}}, \bibinfo {author}
  {\bibfnamefont {S.~A.}\ \bibnamefont {{Levshakov}}}, \bibinfo {author}
  {\bibfnamefont {S.}~\bibnamefont {{Lopez}}}, \bibinfo {author} {\bibfnamefont
  {C.~J.~A.~P.}\ \bibnamefont {{Martins}}}, \bibinfo {author} {\bibfnamefont
  {D.}~\bibnamefont {{Reimers}}}, \ and\ \bibinfo {author} {\bibfnamefont
  {G.}~\bibnamefont {{Vladilo}}},\ }\href {\doibase 10.1093/mnras/stt1356}
  {\bibfield  {journal} {\bibinfo  {journal} {MNRAS}\ }\textbf {\bibinfo
  {volume} {435}},\ \bibinfo {pages} {861} (\bibinfo {year} {2013})},\ \Eprint
  {http://arxiv.org/abs/1307.5864} {arXiv:1307.5864 [astro-ph.CO]} \BibitemShut
  {NoStop}%
\bibitem [{\citenamefont {Ferreira}\ \emph {et~al.}(2014)\citenamefont
  {Ferreira}, \citenamefont {Frigola}, \citenamefont {Martins}, \citenamefont
  {Monteiro},\ and\ \citenamefont {Sol\`a}}]{ferreira1}%
  \BibitemOpen
  \bibfield  {author} {\bibinfo {author} {\bibfnamefont {M.~C.}\ \bibnamefont
  {Ferreira}}, \bibinfo {author} {\bibfnamefont {O.}~\bibnamefont {Frigola}},
  \bibinfo {author} {\bibfnamefont {C.~J. A.~P.}\ \bibnamefont {Martins}},
  \bibinfo {author} {\bibfnamefont {A.~M. R. V.~L.}\ \bibnamefont {Monteiro}},
  \ and\ \bibinfo {author} {\bibfnamefont {J.}~\bibnamefont {Sol\`a}},\ }\href
  {\doibase 10.1103/PhysRevD.89.083011} {\bibfield  {journal} {\bibinfo
  {journal} {Phys. Rev.}\ }\textbf {\bibinfo {volume} {D89}},\ \bibinfo {pages}
  {083011} (\bibinfo {year} {2014})},\ \Eprint {http://arxiv.org/abs/1405.0299}
  {arXiv:1405.0299 [astro-ph.CO]} \BibitemShut {NoStop}%
\bibitem [{\citenamefont {Ferreira}\ and\ \citenamefont
  {Martins}(2015)}]{ferreira2}%
  \BibitemOpen
  \bibfield  {author} {\bibinfo {author} {\bibfnamefont {M.~C.}\ \bibnamefont
  {Ferreira}}\ and\ \bibinfo {author} {\bibfnamefont {C.~J. A.~P.}\
  \bibnamefont {Martins}},\ }\href {\doibase 10.1103/PhysRevD.91.124032}
  {\bibfield  {journal} {\bibinfo  {journal} {Phys. Rev.}\ }\textbf {\bibinfo
  {volume} {D91}},\ \bibinfo {pages} {124032} (\bibinfo {year} {2015})},\
  \Eprint {http://arxiv.org/abs/1506.03550} {arXiv:1506.03550 [astro-ph.CO]}
  \BibitemShut {NoStop}%
\bibitem [{\citenamefont {Levshakov}\ \emph {et~al.}(2002)\citenamefont
  {Levshakov}, \citenamefont {Dessauges-Zavadsky}, \citenamefont {D'Odorico},\
  and\ \citenamefont {Molaro}}]{Deuterium}%
  \BibitemOpen
  \bibfield  {author} {\bibinfo {author} {\bibfnamefont {S.~A.}\ \bibnamefont
  {Levshakov}}, \bibinfo {author} {\bibfnamefont {M.}~\bibnamefont
  {Dessauges-Zavadsky}}, \bibinfo {author} {\bibfnamefont {S.}~\bibnamefont
  {D'Odorico}}, \ and\ \bibinfo {author} {\bibfnamefont {P.}~\bibnamefont
  {Molaro}},\ }\href {\doibase 10.1086/324722} {\bibfield  {journal} {\bibinfo
  {journal} {Astrophys. J.}\ }\textbf {\bibinfo {volume} {565}},\ \bibinfo
  {pages} {696} (\bibinfo {year} {2002})},\ \Eprint
  {http://arxiv.org/abs/astro-ph/0105529} {arXiv:astro-ph/0105529 [astro-ph]}
  \BibitemShut {NoStop}%
\bibitem [{\citenamefont {Carroll}(1998)}]{Carroll}%
  \BibitemOpen
  \bibfield  {author} {\bibinfo {author} {\bibfnamefont {S.~M.}\ \bibnamefont
  {Carroll}},\ }\href {\doibase 10.1103/PhysRevLett.81.3067} {\bibfield
  {journal} {\bibinfo  {journal} {Phys. Rev. Lett.}\ }\textbf {\bibinfo
  {volume} {81}},\ \bibinfo {pages} {3067} (\bibinfo {year} {1998})},\ \Eprint
  {http://arxiv.org/abs/astro-ph/9806099} {arXiv:astro-ph/9806099} \BibitemShut
  {NoStop}%
\bibitem [{\citenamefont {Dvali}\ and\ \citenamefont
  {Zaldarriaga}(2002)}]{Dvali}%
  \BibitemOpen
  \bibfield  {author} {\bibinfo {author} {\bibfnamefont {G.~R.}\ \bibnamefont
  {Dvali}}\ and\ \bibinfo {author} {\bibfnamefont {M.}~\bibnamefont
  {Zaldarriaga}},\ }\href {\doibase 10.1103/PhysRevLett.88.091303} {\bibfield
  {journal} {\bibinfo  {journal} {Phys. Rev. Lett.}\ }\textbf {\bibinfo
  {volume} {88}},\ \bibinfo {pages} {091303} (\bibinfo {year} {2002})},\
  \Eprint {http://arxiv.org/abs/hep-ph/0108217} {arXiv:hep-ph/0108217}
  \BibitemShut {NoStop}%
\bibitem [{\citenamefont {Chiba}\ and\ \citenamefont {Kohri}(2002)}]{Chiba}%
  \BibitemOpen
  \bibfield  {author} {\bibinfo {author} {\bibfnamefont {T.}~\bibnamefont
  {Chiba}}\ and\ \bibinfo {author} {\bibfnamefont {K.}~\bibnamefont {Kohri}},\
  }\href {\doibase 10.1143/PTP.107.631} {\bibfield  {journal} {\bibinfo
  {journal} {Prog. Theor. Phys.}\ }\textbf {\bibinfo {volume} {107}},\ \bibinfo
  {pages} {631} (\bibinfo {year} {2002})},\ \Eprint
  {http://arxiv.org/abs/hep-ph/0111086} {arXiv:hep-ph/0111086} \BibitemShut
  {NoStop}%
\bibitem [{\citenamefont {Chevallier}\ and\ \citenamefont
  {Polarski}(2001)}]{CPL1}%
  \BibitemOpen
  \bibfield  {author} {\bibinfo {author} {\bibfnamefont {M.}~\bibnamefont
  {Chevallier}}\ and\ \bibinfo {author} {\bibfnamefont {D.}~\bibnamefont
  {Polarski}},\ }\href {\doibase 10.1142/S0218271801000822} {\bibfield
  {journal} {\bibinfo  {journal} {Int. J. Mod. Phys.}\ }\textbf {\bibinfo
  {volume} {D10}},\ \bibinfo {pages} {213} (\bibinfo {year} {2001})},\ \Eprint
  {http://arxiv.org/abs/gr-qc/0009008} {arXiv:gr-qc/0009008 [gr-qc]}
  \BibitemShut {NoStop}%
\bibitem [{\citenamefont {Linder}(2003)}]{CPL2}%
  \BibitemOpen
  \bibfield  {author} {\bibinfo {author} {\bibfnamefont {E.~V.}\ \bibnamefont
  {Linder}},\ }\href {\doibase 10.1103/PhysRevLett.90.091301} {\bibfield
  {journal} {\bibinfo  {journal} {Phys. Rev. Lett.}\ }\textbf {\bibinfo
  {volume} {90}},\ \bibinfo {pages} {091301} (\bibinfo {year}
  {2003})}\BibitemShut {NoStop}%
\bibitem [{\citenamefont {Doran}\ and\ \citenamefont {Robbers}(2006)}]{EDE}%
  \BibitemOpen
  \bibfield  {author} {\bibinfo {author} {\bibfnamefont {M.}~\bibnamefont
  {Doran}}\ and\ \bibinfo {author} {\bibfnamefont {G.}~\bibnamefont
  {Robbers}},\ }\href {http://stacks.iop.org/1475-7516/2006/i=06/a=026}
  {\bibfield  {journal} {\bibinfo  {journal} {Journal of Cosmology and
  Astroparticle Physics}\ }\textbf {\bibinfo {volume} {2006}},\ \bibinfo
  {pages} {026} (\bibinfo {year} {2006})}\BibitemShut {NoStop}%
\bibitem [{\citenamefont {{Astier}}\ \emph {et~al.}(2014)\citenamefont
  {{Astier}}, \citenamefont {{Balland}}, \citenamefont {{Brescia}},
  \citenamefont {{Cappellaro}}, \citenamefont {{Carlberg}}, \citenamefont
  {{Cavuoti}}, \citenamefont {{Della Valle}}, \citenamefont {{Gangler}},
  \citenamefont {{Goobar}}, \citenamefont {{Guy}}, \citenamefont {{Hardin}},
  \citenamefont {{Hook}}, \citenamefont {{Kessler}}, \citenamefont {{Kim}},
  \citenamefont {{Linder}}, \citenamefont {{Longo}}, \citenamefont {{Maguire}},
  \citenamefont {{Mannucci}}, \citenamefont {{Mattila}}, \citenamefont
  {{Nichol}}, \citenamefont {{Pain}}, \citenamefont {{Regnault}}, \citenamefont
  {{Spiro}}, \citenamefont {{Sullivan}}, \citenamefont {{Tao}}, \citenamefont
  {{Turatto}}, \citenamefont {{Wang}},\ and\ \citenamefont
  {{Wood-Vasey}}}]{Astier}%
  \BibitemOpen
  \bibfield  {author} {\bibinfo {author} {\bibfnamefont {P.}~\bibnamefont
  {{Astier}}}, \bibinfo {author} {\bibfnamefont {C.}~\bibnamefont {{Balland}}},
  \bibinfo {author} {\bibfnamefont {M.}~\bibnamefont {{Brescia}}}, \bibinfo
  {author} {\bibfnamefont {E.}~\bibnamefont {{Cappellaro}}}, \bibinfo {author}
  {\bibfnamefont {R.~G.}\ \bibnamefont {{Carlberg}}}, \bibinfo {author}
  {\bibfnamefont {S.}~\bibnamefont {{Cavuoti}}}, \bibinfo {author}
  {\bibfnamefont {M.}~\bibnamefont {{Della Valle}}}, \bibinfo {author}
  {\bibfnamefont {E.}~\bibnamefont {{Gangler}}}, \bibinfo {author}
  {\bibfnamefont {A.}~\bibnamefont {{Goobar}}}, \bibinfo {author}
  {\bibfnamefont {J.}~\bibnamefont {{Guy}}}, \bibinfo {author} {\bibfnamefont
  {D.}~\bibnamefont {{Hardin}}}, \bibinfo {author} {\bibfnamefont {I.~M.}\
  \bibnamefont {{Hook}}}, \bibinfo {author} {\bibfnamefont {R.}~\bibnamefont
  {{Kessler}}}, \bibinfo {author} {\bibfnamefont {A.}~\bibnamefont {{Kim}}},
  \bibinfo {author} {\bibfnamefont {E.}~\bibnamefont {{Linder}}}, \bibinfo
  {author} {\bibfnamefont {G.}~\bibnamefont {{Longo}}}, \bibinfo {author}
  {\bibfnamefont {K.}~\bibnamefont {{Maguire}}}, \bibinfo {author}
  {\bibfnamefont {F.}~\bibnamefont {{Mannucci}}}, \bibinfo {author}
  {\bibfnamefont {S.}~\bibnamefont {{Mattila}}}, \bibinfo {author}
  {\bibfnamefont {R.}~\bibnamefont {{Nichol}}}, \bibinfo {author}
  {\bibfnamefont {R.}~\bibnamefont {{Pain}}}, \bibinfo {author} {\bibfnamefont
  {N.}~\bibnamefont {{Regnault}}}, \bibinfo {author} {\bibfnamefont
  {S.}~\bibnamefont {{Spiro}}}, \bibinfo {author} {\bibfnamefont
  {M.}~\bibnamefont {{Sullivan}}}, \bibinfo {author} {\bibfnamefont
  {C.}~\bibnamefont {{Tao}}}, \bibinfo {author} {\bibfnamefont
  {M.}~\bibnamefont {{Turatto}}}, \bibinfo {author} {\bibfnamefont {X.~F.}\
  \bibnamefont {{Wang}}}, \ and\ \bibinfo {author} {\bibfnamefont {W.~M.}\
  \bibnamefont {{Wood-Vasey}}},\ }\href {\doibase 10.1051/0004-6361/201423551}
  {\bibfield  {journal} {\bibinfo  {journal} {A. \& A.}\ }\textbf {\bibinfo
  {volume} {572}},\ \bibinfo {eid} {A80} (\bibinfo {year} {2014})},\ \Eprint
  {http://arxiv.org/abs/1409.8562} {arXiv:1409.8562} \BibitemShut {NoStop}%
\bibitem [{\citenamefont {Albrecht}\ and\ \citenamefont
  {Bernstein}(2007)}]{PCA2}%
  \BibitemOpen
  \bibfield  {author} {\bibinfo {author} {\bibfnamefont {A.}~\bibnamefont
  {Albrecht}}\ and\ \bibinfo {author} {\bibfnamefont {G.}~\bibnamefont
  {Bernstein}},\ }\href {\doibase 10.1103/PhysRevD.75.103003} {\bibfield
  {journal} {\bibinfo  {journal} {Phys. Rev.}\ }\textbf {\bibinfo {volume}
  {D75}},\ \bibinfo {pages} {103003} (\bibinfo {year} {2007})},\ \Eprint
  {http://arxiv.org/abs/astro-ph/0608269} {arXiv:astro-ph/0608269 [astro-ph]}
  \BibitemShut {NoStop}%
\end{thebibliography}%

\appendix

\section{Principal Component Analysis}
 \label{App:AppendixA_PCA}
 
PCA is a nonparametric method that is used in this work in order to constrain the dark energy equation of state $w(z)$. Its performance should not  be compared with parametric methods, since the two are addressing different questions. Instead one should compare it with another nonparametric reconstruction, and this is the approach we follow here. This is useful, for  example, in order to compare  the impact of the different datasets for a certain parametrization.

An advantage of PCA techniques is that they allow
one to infer which and how many parameters can be most
accurately determined with a given experiment. Instead of
assuming a parametrization for the relevant observable (variable) with a set of parameters born of our theoretical prejudices, the PCA method leaves the issue of finding the best parametrization to be decided by the data itself. 

In  \cite{HutererandStarkman} and \cite{PCA2} the PCA approach was applied to the use of supernova data to constrain the dark energy equation of state, $w(z)$. Further work, in \cite{Amendola1}, used this same technique  in combination with fine structure constant measurements. With the goal of making the present article self-contained, here we summarize PCA formalism applied to the use of observables, at different redshifts $(z)$, to constrain the dark energy equation of state, $w(z)$.

 One can divide the relevant redshift range into $N$ bins such that in bin $i$ the equation of state parameter takes the value $w_i$, 
\begin{equation} 
w(z) = \sum_{i = 1}^N w_i \theta_i(z) \,. 
\end{equation} 

Another way of saying this is that $w(z)$ is expanded in the basis $\theta_i$, with $\theta_1 =(1,0,0,...)$, $\theta_2 = (0,1,0,...)$, etc.

In order to find the uncertainty of the parameters $w_i$, we have to build a Fisher information matrix. For that the first step is to  construct the Likelihood function for a generic observable $m(z_i,w_i,c)= 
\mu(z_i, w_i) + c$. For the purposes of this work this can be the apparent magnitude of a supernova, in which case 
\begin{equation}
\mu = 5 \log (H_0 d_L)\,,\qquad c = M +25 - 5 \log H_0
\end{equation}
or it can be the relative variation of $\alpha$ obtained with quasar absorption spectra, for which
\begin{equation}
\mu = \ln[\kappa(\phi-\phi_0)]\,,\qquad c = \ln \zeta\,.
\end{equation}
Then we find
\begin{equation} L(w^i,M) 
\propto \exp \left[ -\frac{1}{2} \sum_{i,j = 1}^N (m-m_F)_i C_{ij}^{-1} (m-m_F)_j 
\right] . 
\end{equation} 
where $m_F$ means $m$ evaluated at the fiducial values of 
the parameters, $m_F = m_F(z_i,w_i^F,c^F)$ and $C^{-1}$ is the inverse of the 
correlation matrix of the data.

Defining $\beta = c-c^F$, and integrating the likelihood in $\beta$, we obtain the 
marginalized likelihood 
\begin{eqnarray} L(\omega_i) \equiv \int_{-\infty}^\infty 
L(\omega_i,\beta) d \beta \nonumber \\
= \sqrt{\frac{2\pi}{A}} \exp\left[ -\frac{1}{2} 
\sum_{i,j = 1}^N (\mu-\mu_F)_i D_{ij}^{-1} (\mu-\mu_F)_j \right] 
\end{eqnarray} %
where $A = \sum_{i,j} C_{i,j}^{-1}$ and 
\begin{equation} D_{ij}^{-1} = C_{ij}^{-1} 
- \frac{1}{A} \sum_{k,l=1}^N C_{kj}^{-1} C_{li}^{-1} . 
\end{equation}
The Fisher matrix can be obtained by approximating  $ L(w_i)$ as a Gaussian in the 
theoretical parameters $w_i$ (the equation of state in each bin) centered around the fiducial model,
and taking the inverse of the resulting correlation function.
The Fisher matrix turns  out to be
\begin{eqnarray} F_{kl} \equiv \left. - 
\frac{\partial^2 \ln L}{\partial w_k \partial w_l}\right|_{w^F} \nonumber = 
\sum_{i,j = 1}^N \left. \frac{\partial \mu(z_i)}{\partial w_k}\right|_{w^F} 
D_{ij}^{-1} \left. \frac{\partial \mu(z_j)}{\partial w_l}\right|_{w^F} , 
\end{eqnarray} 
where the derivatives are evaluated at the fiducial values of the 
parameters.

The uncertainties on the measurement of $w_i$ can be inferred from the Fisher matrix of the parameters $w_i$, specifically from $\sqrt{(F^{-1})_{ii}}$, and they typically increase for larger redshift. One can, however, find a basis in which all the parameters are uncorrelated. This can be done by diagonalizing the Fisher matrix such that $F = W^T \Lambda W$ where $\Lambda$ is diagonal and the rows of $W$ are the eigenvectors $e_i(z)$ or the principal components. These define the new basis in which the new coefficients $\alpha_i$ are uncorrelated and now we can write 
\begin{equation} 
\label{recw} w(z) = \sum_{i = 1}^N \alpha_i e_i(z) \,. 
\end{equation} 

The diagonal elements of $\Lambda$ are the eigenvalues $\lambda_i$ (ordered from largest to smallest) and they define the variance of the new parameters, $\sigma^2(\alpha_i) = 1/\lambda_i$.

\begin{figure}
  \centering
    \includegraphics[width=0.4\textwidth]{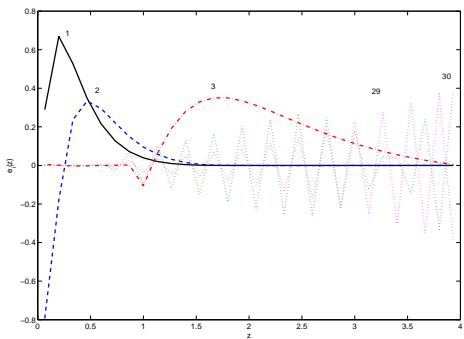}
  \caption[The principal components of $w(z)$]{ The principal components of $w(z)$ for a reconstruction of the 
fiducial model $w(z) = -1$, with supernova type Ia and variation of $\alpha$ measurements. The three best determined and two worst determined eigenvectors are shown and labeled for clarity.}\label{fig:modes}
\end{figure}

In Fig. \ref{fig:modes}, we show the three best determined and two of the worst-determined eigenvectors for $w(z)$ for a reconstruction of the 
fiducial model $w(z) = -1$, with supernova type Ia and variation of $\alpha$ measurements. The best-determined modes peak at relatively low redshifts, while the higher modes (worst determined) have high frequencies and more information at higher redshits.
A way to interpret this parametrization is to realize that the Mth
best-determined eigenvector has precisely $M - 1$ nodes,
leading to the interpretation that the first eigenvector corresponds to the "average of $w(z)$", the
second one to the "first derivative of $w$", the third one
to the second derivative of $w$", etc.

Following ~\cite{HutererandStarkman} and \cite{PCA2} one can now attempt a reconstruction $w(z)$ by keeping only the most accurately determined modes (the ones with largest eigenvalues), and discarding the rest. To do this, we need to decide how many components to keep. We must point out that the weak point of this procedure consists in neglecting the high frequency modes, and typically this will translate into a poorer reconstruction at the highest redshifts. This is one of the reasons why extending the range of the available measurements into the deep matter era is important.

One may argue that the optimal value of modes $M$ to be kept corresponds to the value that minimizes the risk, defined as \cite{HutererandStarkman}
\begin{equation}
risk = bias^2 + variance ,
\end{equation}
with 
\begin{equation} 
bias^2(M) = \sum_{i=1}^N\left( \tilde w(z_i) - w^F(z_i) \right)^2 ,
\end{equation} 
where the notation $\tilde w$ means that the sum in (\ref{recw}) runs from 1 to $M$, and 
\begin{equation} 
variance = \sum_{i=1}^N \sum_{j=1}^M \sigma^2(\alpha_j) 
e_j(z_i). 
\end{equation} 
The bias measures how much the reconstructed equation of state, $w_{\rm
rec}(z)$, differs from the true one by neglecting the high and noisy modes, and
therefore typically decreases as we increase $M$. The variance of $w(z)$,
however, will increase as we increase $M$, since we will be including modes that
are less accurately determined.

An alternative way to decide on the number of optimal modes is to choose the
largest value for which the error is below unity, or equivalently, the RMS
fluctuations of the equation of state parameter in such mode are
\begin{equation}
\langle (1+w(z))^2 \rangle = \sigma_i^2 \lesssim 1\,.
\end{equation}
Having thus determined the optimal number of modes, we proceed with the
normalization of the error following  \cite{PCA2} such that
$\sigma^2 = 1$ for the worst determined mode and normalize the error on the
remaining modes by taking
\begin{equation}
\sigma^2(\alpha_i) \rightarrow \sigma_n^2(\alpha_i) =
\frac{\sigma^2(\alpha_i)}{1+\sigma^2(\alpha_i)} .
\end{equation}
 
	A comparison  of the impact of the two truncation methods (risk method vs. normalization of the error) is presented in  \cite{Amendola1}. The main difference is the effect on the size of the error bars of the reconstruction: the normalization of the error method
appears to give more accurate (closer to the fiducial value) but less precise (more conservative errors) reconstructions when compared with the risk minimization procedure.

\end{document}